\documentclass[12pt]{iopart}

\usepackage{graphics}
\usepackage{graphicx,rotating}

\begin{document}

\topical[Standard Model tests with trapped radioactive atoms]{Standard Model tests with trapped radioactive atoms
\footnote{This is an author-created, un-copyedited version of an article accepted for publication in the Journal of Physics G: Nuclear and Particle Physics. IOP Publishing Ltd is not responsible for any errors or omissions in this version of the manuscript or any version derived from it. The definitive publisher authenticated version is available online at http://dx.doi.org/10.1088/0954-3899/36/3/033101 .}}

\author{J.A. Behr$^1$ and G. Gwinner$^2$}

\address{
$^1$ TRIUMF, 4004 Wesbrook Mall, Vancouver, British Columbia V6T 2A3 Canada}
\address{
$^2$ Department of Physics and Astronomy, University of Manitoba, Winnipeg,
Manitoba  R3T 2N2 Canada}

\eads{\mailto{behr@triumf.ca}, \mailto{gwinner@physics.umanitoba.ca}}

\begin{abstract}
We review the use of laser cooling and trapping for Standard Model 
tests, focusing on trapping of radioactive isotopes.
Experiments 
with neutral atoms trapped using modern laser cooling techniques are 
testing several basic predictions of electroweak unification.
For nuclear $\beta$ decay, 
demonstrated trap techniques include neutrino momentum measurements 
from beta-recoil coincidences, along with methods to 
produce highly polarized samples. 
These techniques have set the best general constraints on non-Standard Model
scalar interactions in the first generation of particles.
They also have the promise to 
test whether parity symmetry is maximally violated,
to search for tensor interactions, 
and to search for new sources of time reversal violation.
There are also possibilites for exotic particle searches.
Measurements of the strength of the weak neutral current can be assisted by
precision atomic experiments using traps loaded with small numbers of radioactive 
atoms, and sensitivity to possible
time-reversal violating electric dipole moments can be improved.
\end{abstract}


%\submitto{\JPG}
\noindent Published in the Journal of Physics G\\
J A Behr and G. Gwinner 2009 J. Phys. G: Nucl. Part. Phys. 36 033101 (38pp)\\
http://dx.doi.org/10.1088/0954-3899/36/3/033101
\maketitle

\section{Introduction}
\label{intro}

This paper will review 
experiments testing symmetries of the Standard Model (SM) of particle physics 
by laser trapping and cooling neutral atoms.
The atom traps make possible new experiments to study an old problem,
nuclear $\beta$ decay.
The laser cooling and trapping techniques
also enable precision atomic measurements,
with the possibility of practical experiments with inherently
small amounts of radioactive isotopes where symmetry-violating 
effects are enhanced. 
Previous reviews of the subject include~\cite{sprouse}.

It is beyond our scope to
cover interesting experiments in weak interactions with ion traps.
Ongoing beta-neutrino ($\beta$-$\nu$) correlation experiments include measuring 
the daughter recoil momentum with a Penning trap~\cite{witch}, 
$\beta^-$-recoil coincidences with a Paul trap~\cite{ganil},
and other neutrino-induced kinematic shifts in a Paul trap~\cite{savard}.

\subsection{The electroweak interaction: what we think we know}
\label{sec-know}
There are several basic features of electroweak unification that 
trap experiments can test.
The photon has ``heavy light'' boson partners W$^{+}$, W$^{-}$, and Z$^0$
which mediate the weak interaction. These are all spin-1 ``vector'' bosons,
which immediately 
implies that the Lorentz transformation properties of the effective 
low-energy four-Fermion contact operators are vector and axial vector.
Measurements using atom traps have constrained 
other interactions 
by improved measurements of the
historically valuable $\beta$-$\nu$ correlation.

For reasons that are not completely understood, the weak interaction is
phenomenologically completely ``chiral'': it only couples to left-handed
neutrinos, and parity is maximally violated. The first 
experiments using the $\beta$-decay of 
laser-cooled polarized atoms have been completed, and 
there
is promise for them to compete with and complement precision measurements of
neutron $\beta$ decay.

The neutral weak coupling of the
Z$^0$ is predicted from the other couplings. At momentum transfer much less
than the Z$^0$ production,  
this has been best tested in
cesium atomic parity violation using thermal atomic beams~\cite{wieman} 
and in intermediate energy M{\o}ller scattering at SLAC~\cite{hughes}.
Atoms with larger atomic number $Z$ have 
larger electron wavefunction overlap with the nucleus, 
enhancing contact interactions
like the weak interaction. For example, atomic parity violation effects, which
measure  the strength of the neutral weak interaction, scale
with $Z^2N$, with additional relativistic enhancements of the electron's
wavefunction and momentum at the nucleus ($N$ is the number of neutrons).  Effects in atoms from 
potential time-reversal violating electric dipole moments are predicted to show similar enhancements,
including enhancements from nuclear structure effects like octupole deformation.
Many of the most promising enhancements at high $Z$ are in elements where
all isotopes are radioactive, inherently limiting the number of atoms that
can be produced. 
The traps enable the possibility of tests in such isotopes,
utilizing precision techniques developed recently in the atomic physics 
community.

The weak couplings are also universal in the sense that the quark couplings
are given in terms of the lepton couplings. The weak coupling between nucleons
is not fully understood, even phenomenologically~\cite{haxton}.
Measurement of the nuclear anapole
moment in traps could resolve a present discrepancy between the anapole moments
of cesium and thallium and low-energy nuclear physics results.

\subsection{How a MOT works }
\begin{figure*} \begin{center}

\includegraphics[width=4in]{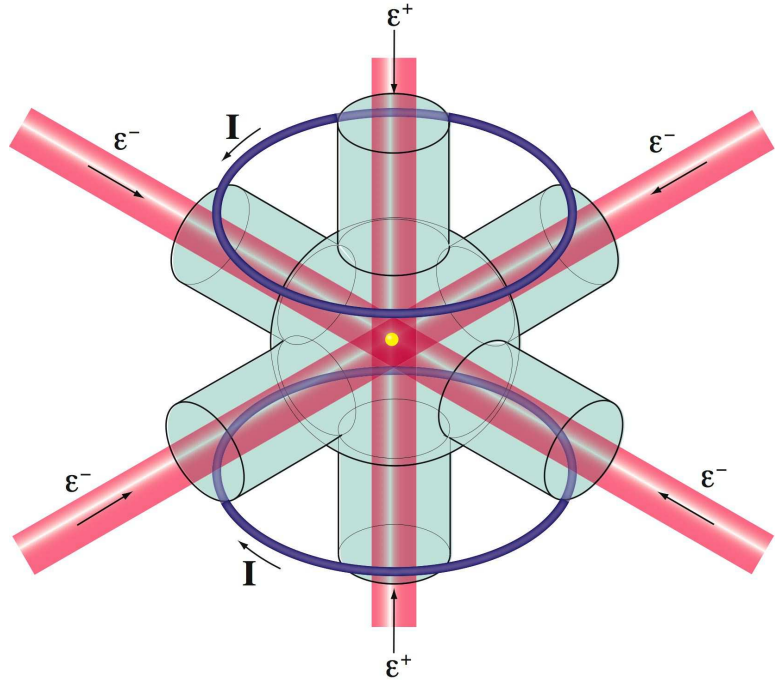}
\end{center}
\caption{
Schematic of a 3-dimensional magneto-optical trap (MOT).
Beams from 6 directions tuned
lower in frequency than atomic resonance slow the atoms.
Coils with opposing currents produce a weak magnetic field
that is linear in all three dimensions,
producing a linear restoring force by increasing the probability of
absorption from whichever beam pushes toward the center (see Figure~\ref{fig:mot1d}).
}
\label{fig-mot}       % Give a unique label
\end{figure*}

It is useful to first 
describe the workhorse trap in this field, the magneto-optical trap
(MOT), sketched in Figure~\ref{fig-mot}.
A MOT can be treated as a damped harmonic oscillator~\cite{raab}. 
The damping is provided by laser light from six directions tuned a few
linewidths lower than
the frequency of an atomic resonance (``to the red'') .
Atoms moving
in any direction see light opposing their motion Doppler shifted closer
to resonance, and preferentially absorb that light and slow down. 
This works naturally in three dimensions to cool the atoms. 

To have a linear restoring force, one must evade the ``optical Earnshaw 
theorem''. 
Consider  Poynting's theorem for the divergence of the 
momentum ${\vec S}$ carried by a plane wave:

\begin{eqnarray}
\noindent {\vec \nabla} \cdot {\vec S} = \frac{c}{4\pi} {\vec \nabla} \cdot ({\vec E}~{\rm x}~{\vec B})
%\indent  
= -{\vec J}\cdot{\vec E} - \frac{\partial u}{\partial t}.
\end{eqnarray}

\noindent This vanishes in a source-free region, when time is averaged over a 
period of the light wave. When the divergence of the Poynting vector is zero, 
there cannot  
be a three-dimensional trap for point particles from 
continuous plane waves of light~\cite{ashkin,Fermi} (by analogy with the Earnshaw
theorem for charged particles in electrostatic fields). 
%For more formal derivations
The loopholes in this theorem are found directly in the conditions listed:
either internal degrees of freedom of the atom are used to make them not
pointlike, or time dependence of the light can be harnessed. 

Ashkin and
Gordon, the authors
of~\cite{ashkin}, immediately  implemented
the use of the `dipole force'
to manipulate the internal
degrees of freedom of the non-pointlike atoms~\cite{ashkinchu}.
If a laser beam is tuned very far off atomic resonance, almost no photons
are absorbed. Instead, the electric field $\vec{E}_{\rm laser}$ 
of the laser light  
polarizes the charge of atoms by the AC Stark shift.
The resulting induced electric dipole $\vec{d}$ 
then couples to $\vec{E}_{\rm laser}$, producing a potential 
energy change of the atom $-\vec{d} \cdot \vec{E}_{\rm laser}$. 
One version of this trap
is simply a laser beam focused to a diffraction-limited spot, which produces
a spatial gradient of $-\vec{d} \cdot \vec{E}_{\rm laser}$ 
in all three dimensions.  
This creates a conservative trap without damping, with typical well depths
of order $10^{-3}$ Kelvin. 
Such dipole force traps are widely used~\cite{dipolereview}. Since they
are conservative, they must always be loaded using other dissipative techniques.

There are also time-dependent forces from pulsed lasers that have been
shown to provide strong forces by alternately exciting and stimulating
photon absorption~\cite{time}. Similar `bichromatic' forces where the
time dependence effectively comes from beats between different light
frequencies have also been harnessed~\cite{soding,metcalf}.

\begin{figure*}
\begin{center}
\includegraphics[width=4in]{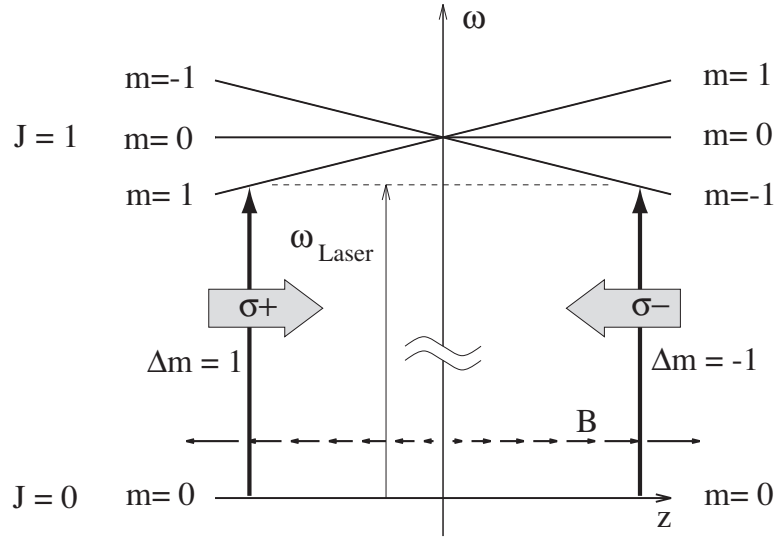}
\end{center}
\caption{Simple 1-dimensional MOT model for an atom with a $J=0$ ground state and $J=1$ excited state. In the center the magnetic field is zero and the laser has a red-detuning of about one to two natural linewidths to provide Doppler cooling. The linear field gradient introduces a Zeeman splitting which together with the handedness  of the counter-propagating beams creates the position dependent force. The MOT quadrupole field produces such linearly changing B fields along each axis,
${\vec B} \approx B_0 (2 {\vec z} - {\vec x}   -  {\vec y} )$.
 In the trapping literature the light polarization is described by the projection of the photons' angular momentum onto a fixed quantization axis (here the z-axis), leading to the shown $\sigma^+ - \sigma^-$ configuration for the MOT; the counter-propagating beams therefore have the same handedness $\epsilon^+$; in addition, the beams in the $x-y$ plane have the same handedness $\epsilon^-$ opposite to that of the $z$ beams (running through the coils in Figure~\ref{fig-mot}). Adapted from~\cite{raab}.
}
\label{fig:mot1d}
\end{figure*}

Dalibard is credited with the most successful idea to manipulate the
internal structure of the atoms to preferentially absorb the beam directing
atoms to the center, the MOT~\cite{dalibard}, which is shown schematically in Figure~\ref{fig-mot}.
The linear part of the restoring force is provided by a weak magnetic
quadrupole field with gradient $\sim$ 10 G/cm 
produced by anti-Helmholtz coils. 
This 
$B$ field changes
sign at the origin, changing the sign of the Zeeman splitting and 
therefore the probability of absorbing circularly polarized light with
opposite handedness in the opposing beams (see Figure~\ref{fig:mot1d}). Coupled with the damping from
the red detuning described above, this makes a dissipative trap that 
cools and confines atoms. The depth can be on order one Kelvin.
The ability of the MOT to cool atoms makes it a typical first trap which
then often feeds conservative traps.  

The MOT's magnetostatic potential is more
than an order of magnitude smaller than what is typically used to confine atoms
directly in magnetostatic traps~\cite{metcalfbergeman}.
The magnetic field of the MOT is mainly 
being used to induce the atoms to absorb one
beam or the other. The result is generally an overdamped harmonic
oscillator, with a cloud of atoms $\sim$1 mm in diameter
collected at the origin. 
Because of the different light polarizations in the six beams,
a normal MOT will have atomic and nuclear spin polarization
close to zero, though modified geometries have been used to deliberately 
spin-polarize atoms~\cite{walker}.
Because of the near-resonant laser light,
MOTs are inherently highly isotope and isomer selective.
The mean lifetime of atoms in the MOT is $\sim$ 1 s at a vacuum of
$10^{-8}$ Torr, limited by the average collision cross-section with background
gas, as the momentum transfer in most collisions is more than adequate to eject
the trapped atom. 
A more complete treatment of laser forces on atoms can be found in~\cite{balykin}.

\subsection{MOT-based tests of the weak interaction}
\label{section-MOTbasedtests}

From the experimental properties of the MOT,
one can immediately see several broad classes of experiments that 
MOTs can assist. 

\begin{figure*} \begin{center}
\resizebox{0.9\textwidth}{!}
{%
  \includegraphics{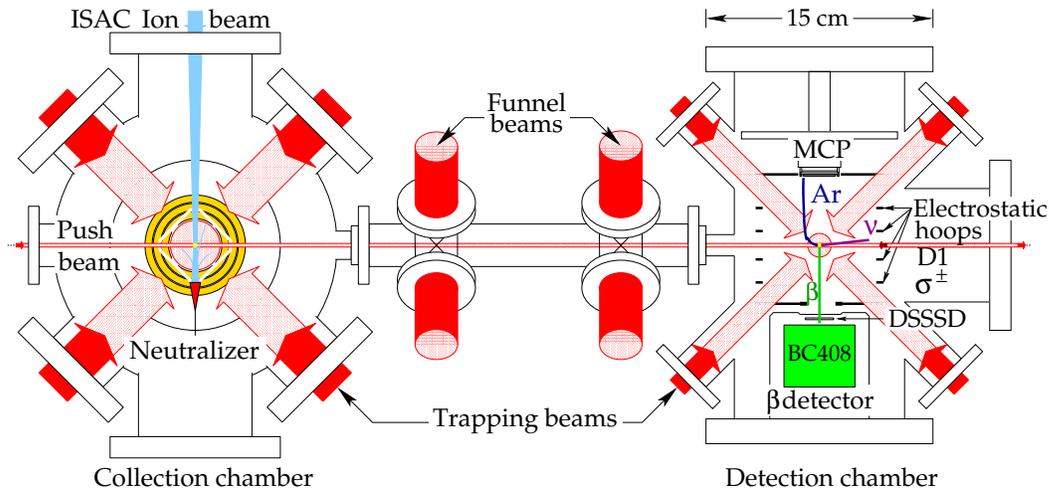}
}
\end{center} 
%\vspace*{5cm}       % Give the correct figure height in cm
\caption{Prototypical
TRIUMF Neutral Atom Trap 2-MOT apparatus. A vapor cell MOT traps radioactives
with 0.1\% efficiency, and then the atoms are transferred with high 
efficiency~\cite{swanson} 
to a second trap with detectors. A uniform electric field
collects ion recoils to a microchannel plate, where their position and
time-of-flight (TOF) with respect to the $\beta^+$ is measured.
An additional beam (``D1 $\sigma^{\pm}$'') can spin-polarize the atoms by
optical pumping when the MOT is off.
}
\label{behrfig1}       % Give a unique label
\end{figure*}

The low-energy ($\sim$ 100 eV) 
nuclear recoils from $\beta$ decay freely escape the
MOT--- they have transmuted to another element so the laser light no
longer matters, and the $B$ field is
very small. Using an apparatus similar to Figure~\ref{behrfig1}, 
the recoils can be accelerated in a known electric field to a microchannel
plate (MCP). Their time and position of arrival at the MCP, along with
their known initial position in the trap cloud (which has size $\sim$ 1 mm),
allows their momentum to be deduced.  
Together with measurement of the $\beta$ 
momentum by more established detection techniques, this 
allows the reconstruction of the $\nu$ momentum
in a much more direct fashion than possible previously. 
(Measurement of the $\beta$ energy is difficult, but there are kinematic
regimes --- recoil momenta less than $Q/c$, where Q is the maximum $\beta$ 
kinetic energy--- for which the neutrino momentum
is uniquely defined from the other kinematic observables~\cite{kofoedhansen};
see Section~\ref{sec-betanu38mK}.) 
Therefore, the angular distribution of $\nu$'s with respect to the $\beta$
direction, the $\beta$-$\nu$ angular correlation, can be measured. 

A variety of methods exist to polarize laser-cooled neutral
atoms and to accurately measure their polarization, 
and some will be described in Section~\ref{sec:polarized}.
Knowledge of the polarization of the
decaying species is a limiting systematic error in many  neutron 
$\beta$ decay and $\mu$ decay experiments. 
For most experimental tests of maximal parity violation, 
the polarization must be known with error less than 0.1\%.

The cold, confined atom cloud also provides a bright source for Doppler-free
precision spectroscopy of high-$Z$ radioactive atoms. 
On the order of 10$^7$ photons/sec are emitted
into 4$\pi$ for a saturated electric dipole transition.
Forbidden transitions that move atoms from one state to another can then
be probed efficiently by laser probes exciting allowed transitions.
The atoms can also be interrogated repeatedly by strong laser, microwave,
and electric fields in well-controlled environments. We will see several
examples of these powerful techniques below.

\subsection{What elements can be trapped?}

\begin{figure*} \begin{center}
%\resizebox{0.75\textwidth}{!}
%{%
  \includegraphics[angle=0,width=0.5\linewidth]{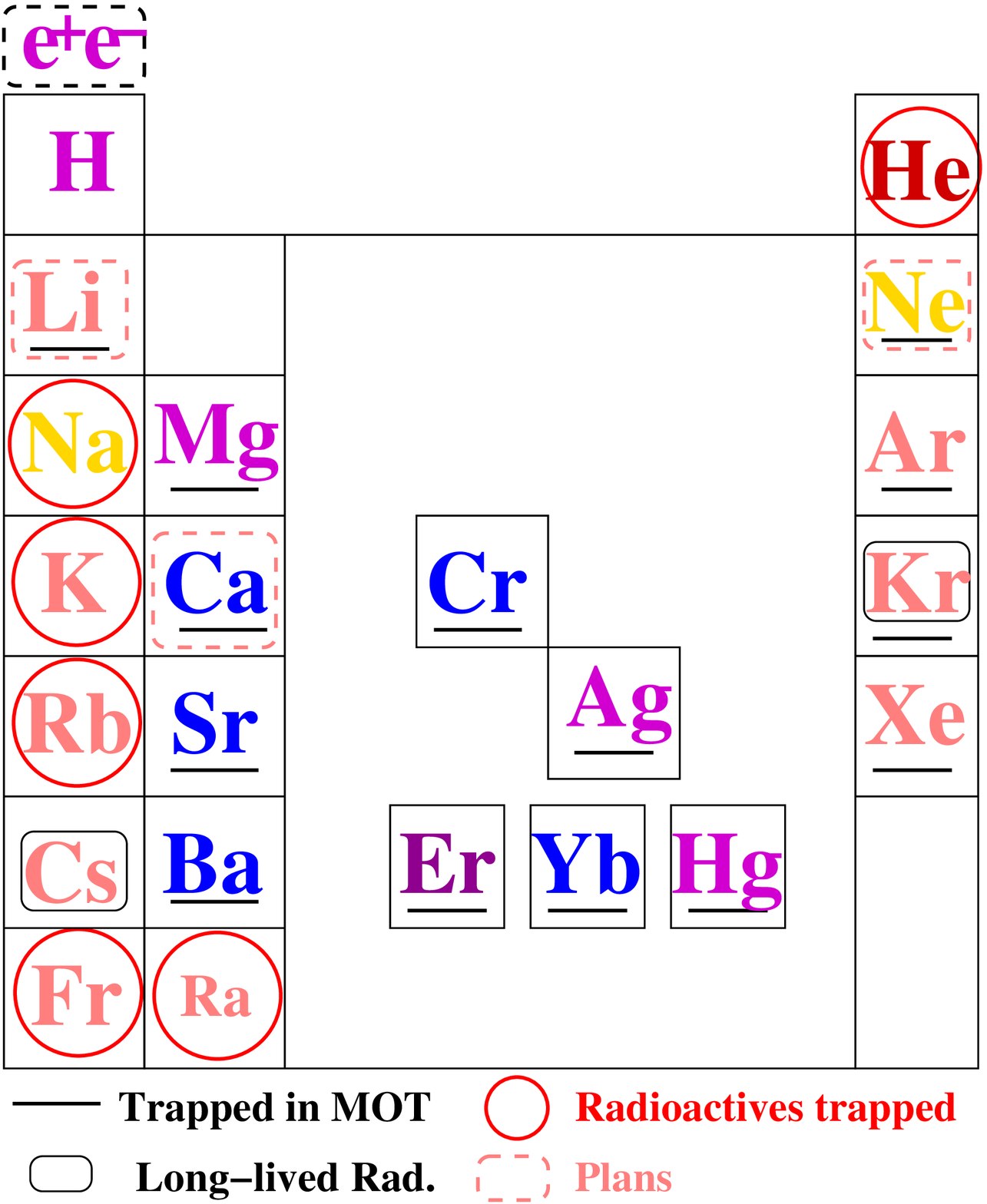}
  \end{center} 
%}
%\vspace*{5cm}       % Give the correct figure height in cm
\caption{
The periodic table of elements that have been laser cooled and/or trapped. 
The approximate
laser wavelength is color-coded (online). Additionally, laser-cooled atoms
include hydrogen, aluminum and iron. 
After G.~D. Sprouse (private communication).
}
\label{behrfig2periodic}       % Give a unique label
\end{figure*}

Tens of thousands of photons must be absorbed to slow atoms from room
temperature, so until recently it was assumed that 
neutral atoms must have reasonably strong cycling transitions
to be trapped (for a cycling or closed transition, spontaneous decay immediately returns the atom back to the state from which it was excited by the laser, leading to continuous re-excitation and strong fluorescence). 
It is best if the excited state atomic angular momentum $J_{\rm excited} = J_{\rm ground}+1$, so that all ground state sublevels can be excited by circularly polarized light.
Figure~\ref{behrfig2periodic} shows elements that have been laser-cooled 
and/or trapped. 

Alkali atoms have a single electron outside a closed noble gas core,
which makes them ideal cases. 
Typically the $s_{1/2} \rightarrow p_{3/2}$ transition is used.
If the nuclear spin $I \neq 0$, then hyperfine splitting produces
two ground states with total angular momentum $F=I \pm \frac{1}{2}$. The transition $F=I+\frac{1}{2} \rightarrow F=I+\frac{3}{2}$ is then cycling,
since decays from $F=I+\frac{3}{2}$ to $F=I-\frac{1}{2}$ do not proceed
by allowed electric dipole transitions. Most of the light is applied just to
the red of this `trapping' transition. Because of weak off-resonant excitation to $F=I+\frac{1}{2}$ excited states,
eventually atoms would accumulate in the $F=I-\frac{1}{2}$ ground state,
so additional, typically much weaker, `repumping' light is also
applied nearly resonant with a transition from that state, feeding atoms back into the right ground state.
Radioactive isotopes of most alkali elements 
(Na, K, Rb, Cs, Fr) have been trapped.

Alkaline earths can also be trapped with 
shorter-wavelength E1 transitions if additional
lasers are used to remove atoms from metastable states.
Barium, which requires complex repumping schemes, has recently been 
trapped~\cite{de}.
Radium atoms have states with potential enhancements of
time-reversal violating electric dipole moment (EDM) and anapole moment 
effects~\cite{flambaum}.
Researchers at Argonne National Lab have succeeded in trapping radium, 
using a mixed transition at 714 nm with $\sim$10\% of an allowed E1 
strength~\cite{guest1,guest2}.

Typically, the first four excited states of a noble gas will have two
metastable states from which there are single-electron cycling transitions
accessible with lasers.
Therefore, they can be trapped if the metastable states are first populated by
some other method, e.g. by using a Penning discharge.
For example, radioactive isotopes of He and Kr have been trapped.
$^6$He and $^8$He have been trapped by the Argonne group, and their
charge radii determined via
optical isotope shift measurements~\cite{luhelium,muellerhelium}. Trace analysis has been done
on long-lived Kr isotopes by the same group~\cite{chen}.

Other elements can be trapped if sufficient effort is given to the
lasers. Laser cooling transitions in a number of unusual species
were proposed by Shimizu~\cite{shimizu}.
Reference~\cite{adams} reviews the successfully trapped elements.
Since that review, 
stable isotopes of Ag~\cite{walther}, Cr~\cite{bell}, Yb~\cite{honda}, 
and Hg~\cite{hachisu} 
have been trapped in a MOT for studies of clock standards,
Fermi degenerate gases and Bose condensates, 
and EDM searches. These elements still have relatively
simple electronic structures.

In contrast, 
the complex rare earth atom erbium has now been laser-cooled and
trapped~\cite{mcclelland,berglund}.
Erbium has an optical transition possibly useful for a time standard.
A single-frequency laser was used.
There are more than 50 states between the excited state in the main laser 
transition and the ground state.
The large atomic angular momentum
($J=7$ ground state) makes transitions to most of these states weak, 
minimizing the loss of population to long-lived low-$J$ metastable states.
The high $J$ also produces a large atomic magnetic moment, 
so the weak MOT quadrupole field is 
thought to help contain the atoms during the time they are in metastable
states.
This opens the door to trap other complex systems with large atomic angular
momentum, for whatever specific cases prove to be useful.

A rather more exotic possibility would be the trapping of orthopositronium,
for which work has been done at Tokyo Metropolitan University~\cite{positronium}.
The goal of this would be a Bose-Einstein condensate to produce coherent
annihilation $\gamma$-ray emission~\cite{amills}. 

\subsubsection{ Loading the shallow MOT for alkalis}

The MOT depth is on order one Kelvin. Each group trapping
radioactive isotopes has invested large efforts to learn to load a MOT 
efficiently in geometries appropriate for their particular experiments. 
The creation of the radioactives inherently involves a production target
where nuclear reactions produce too much background for any type of
experiment, so the isotopes must be transported away from this region.
The final vacuum, desired at the 10$^{-10}$ Torr level, also involves challenges.
The first two groups which trapped radioactives solved these problems
in quite different ways, and most efforts since then have made improvements
along the same lines. 

In the initial Stony Brook work trapping 
$^{79}$Rb~\cite{gwinner}, a heavy ion beam induced fusion reactions in 
a foil that was a combination target and surface ionizer.
The products were
transported as a low-energy ion beam~\cite{behrnim}. The ions were implanted
in a surface with a low work function (yttrium) to keep the evolving atoms
neutral. The atoms were collimated into an atomic beam feeding a 
vapor-cell lined with silicone polymer coatings to which the alkali atoms
do not stick~\cite{swenson,fedchak}. 
The vapor cell confined the atoms for 
many passes through the beams and many chances to be 
trapped, after re-thermalization upon contact with the walls replenished
the low-velocity tail of the Maxwell-Boltzmann velocity distribution~\cite{holbrow}.

The short-lived isotope $^{21}$Na trapped at Berkeley was produced as a 
collimated atomic
beam from a hot magnesium oxide production target. The atoms were 
slowed longitudinally by an unopposed laser beam 
as they traversed inside a tapered solenoid
utilizing the Zeeman effect to keep the atoms in resonance (a 
`Zeeman slower') before they entered the trap~\cite{lushang}. 

We show the TRIUMF Neutral Atom Trap (TRINAT) 
system as a typical example for the loading and preparation 
process~\cite{gorelov}
(Figure~\ref{behrfig1}).
A mass-separated 30 keV ion beam from TRIUMF's ISAC facility~\cite{dombsky} 
is stopped in a 900$^{\rm o}$C Zr foil, adapting a geometry pioneered at 
Los Alamos~\cite{guckert}
to use a conical neutralizer to minimize implantation depth.
Only atoms moving at less than about $5-10$\% of room temperature velocities
can be trapped.  
Typical efficiencies for trapping
atoms in a vapor cell are 0.1 to 1\%.
To avoid the large radioactive background from untrapped atoms, both in the 
untrapped vapor and on the walls, 75\% of the cooled atoms are transferred
in TRINAT to a second MOT where the experiment takes place.
The transfer time is 25 msec. Details of the transfer process are in~\cite{swanson}.  
For atomic physics experiments on stable species, 
the two-MOT arrangement is also common
to improve the vacuum, avoid the small backgrounds from the
Doppler-broadened vapor, and allow specialized apparatus surrounding
the second MOT.

Vapor-cell MOT efficiencies of ~50\% have been reported in stable
species~\cite{stephens}, where efficiency is defined as the percentage
of incoming atoms that are loaded into the trap. 
Efficiencies for radioactive species have not  
exceeded $5 \times 10^{-3}$, reported by the Los Alamos 
group~\cite{guckert,dirosa}.
Possible ways to increase the capture velocity that have
been used on stable species include 
frequency combs farther to the red~\cite{kasevich}, white light
slowing~\cite{rasel}, and
light-assisted desorption~\cite{calabrese2}.
The MOT relies on the emission process being front-back symmetric with
respect to absorption, so it is limited by the spontaneous atomic decay
rate, and it does not help to increase laser power beyond saturation.
Stimulated forces, like the bichromatic forces mentioned above~\cite{metcalf}, 
have been demonstrated to slow
stable Cs in one dimension 
from room temperature over a distance of 10 cm~\cite{soding}, and they
have the promise of being limited only by the laser power applied (admittedly, for many atoms lasers have insufficient power to 
saturate the spontaneous forces.)

The traps are highly element, isotope, and isomer selective. For example, at TRIUMF/ISAC
the mass-separated 
$A=38$ ion beam has 20 times more of the ground state of the isotope
$^{38}$K (spin and parity $I^\pi$=3$^+$, t$_{1/2}$$\sim$7 minute)
than the nuclear isomer $^{38{\rm m}}$K of interest 
($I^\pi$=0$^+$ t$_{1/2}$$\sim$1 sec). 
The $I^\pi$=3$^+$ nuclear ground state 
has an atomic ground state hyperfine splitting of  
1.4 GHz, and these states straddle equally the location in atomic energy of the
0$^{+}$ isomer. The MOT works at frequencies from 
5 to 50 MHz to the red of resonance, 
but more importantly, 
two frequencies are required to trap the hyperfine-split
3$^+$ state. Hence, the one frequency applied to the MOT only traps the 0$^+$ 
nuclear isomer. 
No recoil-$\beta$ coincidences were observed from the decay of the 
3$^+$ state~\cite{gorelov}.

\section{Atom Traps for Decay Experiments}
\label{decayexperiments}

In this section we describe the use of atom traps for nuclear beta decay, both
with spin polarization and without. We also describe missing-momentum
searches for sterile neutrinos in $\beta$ decay and electron capture,
along with other exotic particle searches in isomer decay.

We expand informally 
upon our introduction to $\beta$ decay in Section~\ref{sec-know}. 
For a more complete and technical review, see~\cite{severijns}.
At these low momentum transfers, $\beta^-$ decay in 
the Standard Model or any extension based on
exchange of very massive bosons reduces to a sum of 
4-fermion contact interactions~\cite{leeyang}:

\begin{eqnarray}
H_{\rm int} & = &
\sum_X
(\bar{\psi}_p O_X \psi_n)(C_X \bar{\psi}_e O_X \psi_\nu + C_X' \bar{\psi}_e O_X \gamma_5 \psi_\nu)
\label{eq-H}
\end{eqnarray}

\noindent 
where $O_X$ denotes operators with the five different possible Lorentz 
transformation properties $X$ ---
vector (V), axial vector (A), tensor (T), scalar (S), and pseudoscalar (P) ---
and implicitly includes all necessary contracted relativistic 4-indices.
This interaction then is invariant under Lorentz
transformations. 

The combinations of $C_X$ and $C_X'$ produce projection operators 
$1\pm \gamma_5$
which project out either left or right-handed neutrinos.
In the Standard Model, the interaction between quarks and leptons is `$V-A$', 
so if we had written the interaction between quarks and leptons, then
$C_V = C_V'$ and $C_A = -C_A'$, 
the combination given by exchange of the spin-1 W boson. 
Then only left-handed neutrinos are emitted.

The absolute value of $C_A$ departs from unity as QCD combines quarks 
into nucleons, but the interaction still produces only left-handed neutrinos. 
Similarly, though all other constants besides V and A would be zero in the 
SM quark-lepton interaction, similar terms become in principle 
allowed again in the nucleon-lepton interaction; 
these are termed ``induced (by QCD) currents''. 
Many of the induced currents would violate what is called
G-parity, which reduces to charge symmetry in the first 
generation of particles,
and were therefore termed ``second-class currents'' by 
Weinberg~\cite{weinberg} and removed from the SM.
Thus the scalar constants $C_S$ and $C_S'$ are still zero in the SM, as they
are 2nd-class currents, and they also violate the conserved vector current 
(CVC)
hypothesis. In fact fundamental quark-lepton scalars cannot be distinguished
experimentally from induced scalar interactions~\cite{ormandbrownholstein}.
Greater care must be taken to distinguish experimentally between 
tensor non-SM quark-lepton interactions and allowed induced tensor currents.
In isobaric analog decays (like that of the neutron, or $^{21}$Na and
$^{37}$K below) 
the extra induced tensor-order
interactions are either given by CVC in terms of the electromagnetic moments, 
or vanish because they are 2nd-class currents~\cite{holstein}.

The general expression for the nuclear beta-decay rate $W$ in terms of the 
angular correlations and distributions of the 
leptons, including the possible spin-polarization of
the nucleus, is given (using lepton momenta ${\vec p}$ and energy
$E$ and 
nuclear spin polarization and unit direction ${\vec I}$ and ${\hat i}$) 
by~\cite{jtw}:

\begin{eqnarray}
\nonumber W dE_e d\Omega_e d\Omega_\nu  =
\frac{F(\pm Z, E_e)}{(2\pi)^5} 
p_e E_e (E_0-E_e)^2 dE_e d\Omega_e d\Omega_\nu \frac{1}{2} \xi ~~~~~\\
\nonumber ~~~~~\left[ 1 + a \frac{{\vec p_e} \cdot {\vec p_\nu}}{E_e E_\nu} 
+ b\frac{m}{E_e}
+c\left( \frac{1}{3}\frac{{\vec p_e} \cdot {\vec p_\nu}}{E_e E_\nu} - 
\frac{{\vec p_e}\cdot {\hat i}}{E_e E_\nu} \right)
\left(\frac{I(I+1)-3\langle{(\vec I}\cdot \hat{i})^2\rangle}{I(2I-1)}\right)
\right. 
\\
~~~~~\left. + \frac{\langle {\vec I}\rangle}{I} \cdot \left(A_\beta \frac{{\vec p_e}}{E_e} + 
B_\nu \frac{{\vec p_\nu}}{E_\nu} + 
D \frac{{\vec p_e} \times {\vec p_\nu}}{E_e E_\nu} \right) \right] ,
\label{eq-W}
\end{eqnarray}
where $F$ is the Fermi function.
We will discuss below trap 
measurements of the $\beta$-$\nu$ correlation coefficient
$a$, the $\beta$ asymmetry with respect to spin $A_{\beta}$, the 
$\nu$ asymmetry with respect to spin $B_{\nu}$, and the time-reversal violating
correlation coefficient $D$. 
The 2nd-rank tensor alignment term with coefficient $c$ occurs for
nuclear spin $I \geq 1$.
Explicit expressions for these experimental observables as a function of the
$C_X$ constants were worked out in~\cite{jtw}, and can be found rewritten in 
explicitly chiral notation in the review of~\cite{severijns}. Rather than
rewrite these expressions, we discuss qualitative features here.
We ignore here observables that measure the spin-polarization of the leptons,
as these have not been pursued as yet with traps.

In Eq.~(\ref{eq-W}) we can see that
the correlations are all normalized by the
change in decay strength due to the term $b$.
The decay rate and angular distributions are given by the absolute square of
the matrix elements of $H_{\rm int}$.
That produces cross-terms between
new interactions and the SM interactions that are therefore linear in the
small new coupling coefficients.
Such `Fierz interference terms', collected together as $b$ in 
Eq.~(\ref{eq-W}), always produce left-handed neutrinos, just
as the SM does. So searches confined to them already assume the complete
chirality and good time reversal symmetry of the SM. This is a natural thing to do in many theories,
and many limits from particle physics in the literature 
simply assume this chirality without qualification.

Terms are also produced that are squares of the new interactions. We will
give simple arguments below why the beta-neutrino correlation is sensitive
to these. These terms are more general in the sense that they are 
insensitive to the chirality and 
time-reversal symmetry properties of the new interactions.

To take an example of an explicit model,
one possible source of non-Standard model scalar and tensor interactions is
supersymmetry. Reference \cite{profumo} has shown in a wide variety of SUSY 
models that left-right mixing between supersymmetric partners of the 
first-generation fermions 
can generate terms as large as 0.001 in the Fierz interference scalar-vector 
and tensor-axial vector terms. This left-right sfermion mixing
is difficult to constrain in particle physics searches.
It is a goal for many of the correlation experiments discussed below to reach such
sensitivity. 

We also note here 
one possibly confusing fact: a spin-0 leptoquark (i.e. a particle
explicitly changing leptons into quarks) can generate both 4-fermion scalar 
and tensor effective interactions~\cite{herczeg}. Consequently, a fundamental interaction producing a 4-fermion effective tensor interaction does not imply some
very exotic spin-2 particle.

\subsection{Recoil momentum from traps: Beta-neutrino correlations, motivation}

Historically, the $\beta$-$\nu$ angular correlation 
(the $a$ term in Eq.~(\ref{eq-W}))
has provided some of
the best evidence that the effective contact interaction was primarily 
vector and axial vector, 
which in modern theories is due to exchange of the spin-1 bosons.

Adelberger pointed out the utility of such measurements in 
pure Fermi decay to constrain scalar interactions~\cite{adelberger1992}. 
One can make a simple helicity argument to show that $a$=1 for these 
decays. The leptons
are produced with opposite helicity in the Standard Model interactions. 
For $I^\pi$ = 0$^+$$\rightarrow$0$^+$ decays, where the leptons must
carry off no angular momentum, they cannot be emitted back-to-back.
Thus these experiments
are insensitive to the absolute chirality of the couplings, and only depend on 
the relative helicity of the two leptons. 
They are sensitive to the sums
of absolute squares of the scalar or tensor interactions in Eq.~(\ref{eq-H}).
Although one is measuring the square of small terms,
this is a large advantage when looking for wrong-chirality interactions that do
not interfere with the Standard Model's one-sign chirality.

The Fierz interference 
terms also modify the decay rate and 
therefore the
normalization of the angular distribution.  The $\beta$-$\nu$ correlation 
also has linear sensitivity
to some of the new physics. 

\subsection{Beta-neutrino correlations: experiment}
\subsubsection{$\beta$-$\nu$ correlation of $^{38{\rm m}}$K}
\label{sec-betanu38mK}

TRIUMF's Neutral Atom Trap Group (TRINAT) 
has published its $\beta$-$\nu$ correlation 
result for $^{38{\rm m}}$K, a pure Fermi decay sensitive to scalar 
interactions~\cite{gorelov}. The result for the angular distribution
coefficient 
$a$=0.9981 $\pm$ 0.0030 (statistical) $\pm$ 0.0037 (systematic)
is in agreement with the
Standard Model value of unity. 
It has somewhat greater accuracy than the Seattle/Notre Dame/ISOLDE work 
in $\beta$-delayed proton decay of $^{32}$Ar~\cite{adelberger}, which set
the previous best general limits on scalars coupling to the first generation
of particles. (The $^{32}$Ar work, 
re-evaluated after re-measurement of the decay energy, 
gives $a$=0.9980$\pm$0.0051 (statistical) with systematic error to
be determined~\cite{garcia}.)
The TRINAT work 
was done with two thousand atoms trapped at a time, at densities 
less than 0.5\% of those in the Berkeley work, avoiding the possibility
of trap density distortions (see below).

\begin{figure*} \begin{center}
%\resizebox{0.75\textwidth}{!}
{%
  \includegraphics[width=5in]{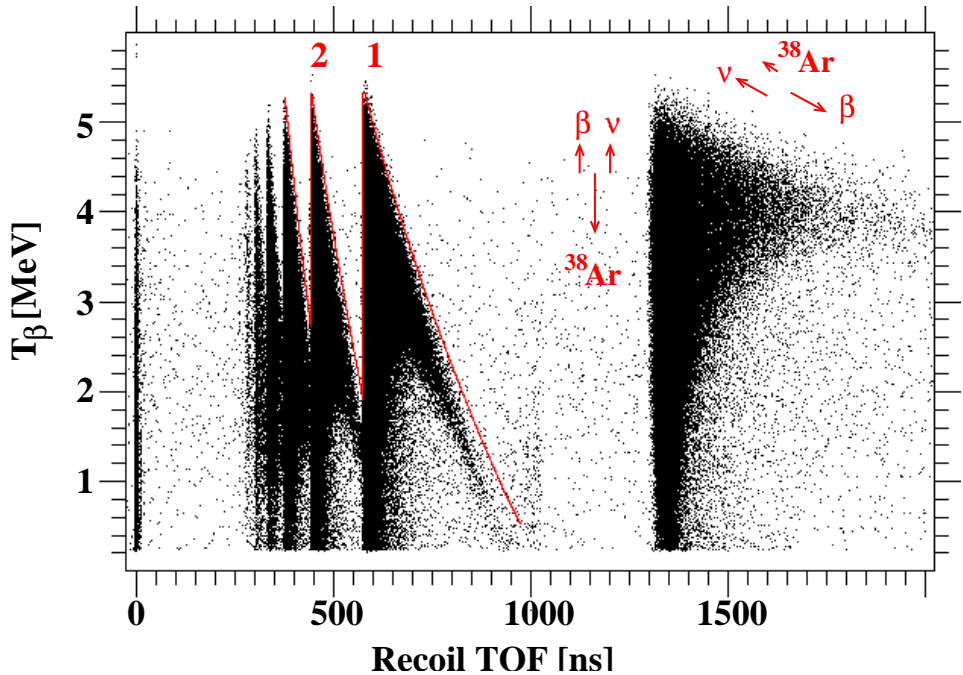}
}
\end{center} 
%\vspace*{5cm}       % Give the correct figure height in cm
\caption{
Scatter plot of $^{38{\rm m}}$K $\beta$-recoil correlation data. 
The recoils are produced in several charge states ranging from neutral
atoms to ions of charge one through six, which
are separated by their time-of-flight (TOF)
in the uniform electric field of Fig.~\ref{behrfig1}. The arrows denote
lepton and recoil momenta (see text). 
%Lowest of 16 $\beta$ kinetic energy $T_{\beta}$ 
%bins for the theory fit to the Ar$^{+1,+2,+3}$ data
%in coincidence with the $\beta$ for $^{38{\rm m}}$K $\beta$-$\nu$ correlation.
%The dip in Ar$^{+1}$ is from the finite MCP size: the dashed curve shows the
%theory result for a larger MCP that would collect all ions; thus the 
%acceptance of the MCP and the electric field must be well-understood~\cite{gorelov}.
}
\label{fig:betanu}       % Give a unique label
\end{figure*}

The nuclear detection is done using the right-hand MOT apparatus of 
Figure~\ref{behrfig1}, as discussed in Section~\ref{section-MOTbasedtests}.
A scatter plot of 500,000 $\beta^+$-recoil coincidence events 
from the decay of
$^{38{\rm m}}$K is shown in Figure~\ref{fig:betanu}.
The solid lines are the kinematic loci that would result from back-to-back
pointlike detectors. If the two leptons are emitted in a similar direction, 
the recoil momentum is very similar (the $\beta$'s are relativistic so the
energy sharing between leptons matters little), producing large numbers of
events with similar time-of-flight (TOF) for each charge state. 
When the leptons are emitted  close to back-to-back, the recoil momenta
are much smaller, producing the arcs at longer TOF. 
One can immediately see qualitatively that the leptons are rarely emitted
back-to-back. These spectra are binned in TOF and $\beta$ energy, and a 
fit to a detailed Monte Carlo simulation is used to extract the quantitative
angular distribution~\cite{gorelov}.

Background events can also be rejected if they are not kinematically allowed.
At 1025 ns in TOF in Figure~\ref{fig:betanu}, there are events at part
per thousand probability at low
measured $\beta^+$ energy. These originate from $\beta's$ that  are emitted
towards the MCP-- sending recoils away from the ion micro-channel plate (MCP) to be
collected by the electric field at later times-- then scatter off
material and into the $\beta^+$ detector.

\begin{figure*} \begin{center}
{
\includegraphics[angle=90,width=3.0in]{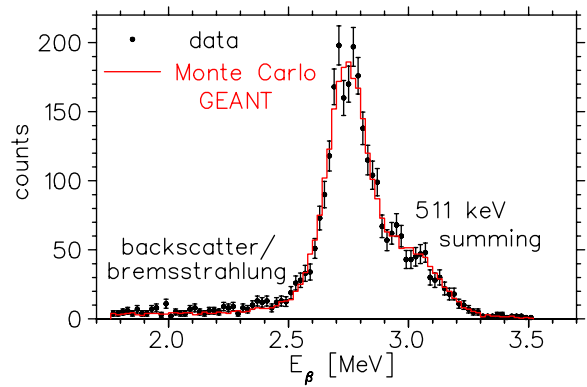}
}
\end{center} 
\caption{
Energy response of the $\beta^+$ detector plastic scintillator of 
Figure~\ref{behrfig1} to `monoenergetic' $\beta^+$'s
(with total energy from 2.7-2.8 MeV)
as determined from the $\beta$-recoil angle and recoil momentum (see text).
Events where less
energy is deposited in the detector are produced by backscattering out of
the detector and by emission of bremsstrahlung $\gamma$-rays. Some annihilation
511 keV $\gamma$-rays Compton scatter in the scintillator, producing the
higher-energy knee.  
The width of the main peak is dominated by the energy resolution
of the plastic scintillator. 
}
\label{lineshape}
\end{figure*}

The energy response of the $\beta$ detector 
is critical in this type of 
$\beta$-recoil coincidence measurement.  
TRINAT can determine detector response functions in situ from the actual data.
This is typically done in high-energy experiments but never before
for low-energy $\beta$ decay.
Consider, for example, a TOF cut of 750-850 ns in Figure~\ref{fig:betanu}.
These ions of charge state +1 come from events with 
a narrow range of $\beta^+$ kinetic energy centered around 2.2 MeV. 
Events with lower detected $\beta^+$ energy 
are produced by the imperfect $\beta^+$ energy determination
of the $\beta^+$ detector in Figure~\ref{behrfig1}. The $\beta^+$
energy can be more precisely reconstructed by 
including the $\beta^+$ direction 
information from the position-sensitive $\Delta$E detector. 
%Figure~\ref{lineshape} 
%shows this example of the 
%energy response of the $\beta$ detector to the resulting monoenergetic
%$\beta$'s.
Figure~\ref{lineshape} 
shows the energy response of the $\beta$ detector to the
monoenergetic $\beta$'s obtained this way.

\paragraph{Limits on scalar couplings}

\begin{figure*} \begin{center}
{
\includegraphics[angle=90,width=3.0in]{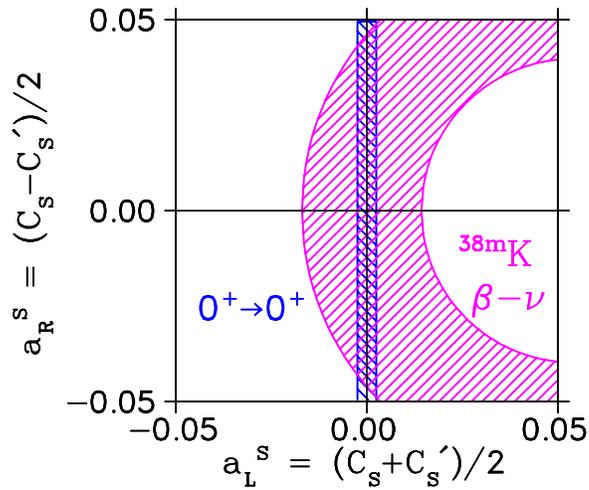}
}
\end{center}
\caption{
Constraints at 90\% CL on scalar interactions, from 
0$^+$ to 0$^+$ $ft$ values~\cite{towner} (rectangle) and from 
the TRINAT $^{38{\rm m}}$K $\beta$-$\nu$ correlation~\cite{gorelov} (concentric
circles).
}
\label{fig-scalar}
 \end{figure*}

The limits on scalar interactions from two sources are shown in 
Figure~\ref{fig-scalar}. There are tight constraints on the scalar-vector
Fierz interference term from the superallowed $ft$ values as a function of
energy release~\cite{towner}, 
because the Fierz term depends on the $\beta$ energy 
(Eq.~(\ref{eq-W})). 
The $\beta$-$\nu$ correlation sets more general constraints on scalars that
couple to either left or right-handed neutrinos~\cite{adelberger}.
The $\beta$-$\nu$ correlation results
from $^{32}$Ar have the same centroid as $^{38{\rm m}}$K with somewhat
larger total error~\cite{garcia}, and when that
final error is decided the allowed area will decrease somewhat.
Powerful but model-dependent constraints from 
$\pi$$\rightarrow$$e \nu$ decay are 
considered in~\cite{campbell}. 
A scalar interaction coupling to right-handed neutrinos produces a 
mass for the SM neutrino, and order-of-magnitude estimates for this effect
were done in~\cite{ito}.

\subsubsection{$\beta$-$\nu$ correlation of $^{21}$Na}
The laser-trapping group of Lawrence Berkeley Lab
had earlier 
published the first $\beta$-$\nu$ correlation using an atom
trap~\cite{scielzolbl}. 
Their abstract quotes the result   
$a$=0.5243$\pm$ 0.0091 for $^{21}$Na, which has a Standard Model prediction 
of 0.558.  
A Gamow-Teller branch to an excited state was subsequently remeasured by
several groups,
although to explain the full deviation the branch would have had to be
7\% rather than the compiled value of 5.0$\pm$0.13\%, and the new more
precise measurements produced only slight changes from the
compiled value~\cite{hardyna21}.
 
The Berkeley group 
presented evidence for a dependence of $a$ on the density of atoms
trapped for more than $10^5$ atoms trapped. 
If an extrapolation to zero density was done,   
the value for $a$ was brought into agreement with the Standard Model,
$a$=0.551$\pm$0.0013$\pm$0.006~\cite{scielzolbl}. 
They have since definitively 
characterized the effect  
(see the following Section~\ref{molecules}).

\subsubsection{Trap-produced perturbations}

Since the atoms trapped are not ideal point particles, 
it is important to note some of the complications produced by 
atomic physics and trap effects.

\paragraph{Formation of ultracold molecules}

\label{molecules}

Deliberate formation of ultracold molecules has produced a large number of
precise experiments in stable species. Electric dipole transition matrix
elements between $p$-state and $s$-state atoms can be deduced by the frequency
dependence of photoassisted collisions~\cite{reviewphotoassisted}.
The ultracold 
molecules themselves are very interesting to chemists and to proposed
precision measurements including anapole moments,  searches for permanent electric dipole moments,
and quantum computation~\cite{demillequantumcomputation}.
Unfortunately, they also produce malevolent effects
in beta-neutrino correlations. 

The Berkeley group
suggested a possible mechanism for a density-dependent effect in their
measurement of the $\beta$-$\nu$ correlation. 
Distortions of the recoil momentum are 
produced when the decay originates from a 
molecular dimer magnetostatically confined within the MOT's weak 
quadrupole B field.
They have observed the molecular dimers, and inhibited their formation
by using a dark spot MOT.

They also developed a high-statistics data technique, measuring the 
shakeoff electrons with high efficiency in coincidence with the recoils.
The resulting TOF spectrum is almost equivalent to a momentum spectrum of
the recoils. Sensitivity to $a$ is inherently lower than in the $\beta$-recoil
coincidence, but the overwhelmingly higher count rates produce a smaller 
statistical error on $a$.
After accounting for the measured dependence of the measured $a$ on 
trap density, the 
result is
$a$=0.5502$\pm$0.0060~\cite{vetterprc}, in agreement with the 
Standard Model
prediction 0.553$\pm$0.002.

It is important to realize that the frequencies and strengths of the 
photoassociation resonances are a strong function of fine and hyperfine
structure, and their effects on the determination of $a$ 
must be determined in future experiments on an isotope-by-isotope basis.

\paragraph{Doppler shifts are small}

The remaining Doppler shifts after laser cooling 
are negligible for nuclear $\beta$-$\nu$ angular
correlation decay. Possible experiments in electron capture producing
recoils with kinetic energies $\sim$ eV will 
require sub-Doppler cooling (see Section~\ref{sterile} below), which 
generally comes for free in careful MOT experiments~\cite{walhout}.

\subsubsection{Atomic charge state dependence on recoil momentum}
Work at Berkeley and TRIUMF
has confronted an additional systematic error common to
most other recoil momentum measurements, the possibility that the final atomic
charge state depends on recoil momentum~\cite{scielzo2}. 
If the charge state of the atom 
depends on recoil or beta momentum, the deduced angular correlations are
perturbed. 
This effect is a potentially important correction to many atom and ion trap 
$\beta$-$\nu$ experiments,
so we sketch some details here. 

Momentum-dependent shakeoff was first postulated, modelled, 
and measured in $^6$He $\beta^-$ decay work at Oak Ridge~\cite{carlsonhe6}. 
Atomic electrons in the daughter can be 
treated as suddenly moving with the recoil
velocity. A plane wave expansion of the resulting sudden approximation
matrix element produces an effect
proportional to the square of the recoil velocity. 
Hence, the sudden approximation to lowest order produces 
a distortion of the recoil energy spectrum of the form
$(1+s E_{\rm rec}/E_{\rm max})$, where $E_{\rm rec}$ is the recoil kinetic
energy, and $E_{\rm max}$ is the maximum value of the recoil kinetic energy.
A simple estimate by the Berkeley group 
relates the size of the parameter $s$ to atomic dipole oscillator strengths,
and suggests that it could be larger in 
$\beta^+$ decay~\cite{scielzo2} because of the difference in atomic
binding energies. 

\begin{figure} \begin{center}
\resizebox{0.45\textwidth}{!}{%
\includegraphics{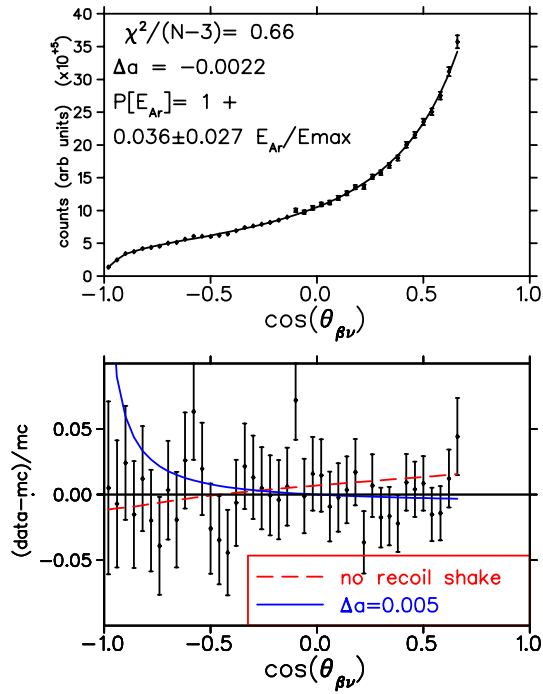}
}
\end{center} 
\caption{
Top: The distribution of detected coincidence events as a function
of the angle between $\beta^+$ and neutrino in $^{38{\rm m}}$K decay, 
as determined in the TRINAT apparatus. 
If all recoils were collected, this would be a straight line
determining $a$ (see Eq.~(\ref{eq-W})); here a Monte Carlo simulation
includes the detector acceptance. 
Bottom: Difference between experiment and model, normalized to model.
%Using TRINAT's reconstructed $\beta$-$\nu$ angular distribution data,
The dependence of recoil electron shakeoff on the recoil momentum produces
a different effect on the angular distribution than changes in $a$, so it
can be simultaneously fit and shown to be small (see text).
}
\label{behrfig2}
\end{figure}

In the absence of detailed calculations of the momentum-dependent shakeoff,
it can be constrained by fitting the above expression to experimental data
and letting $s$ float along with the coefficient $a$ of the $\beta$-$\nu$
angular distribution, because they have different dependence on the kinematic
variables. This was done in two separate analyses with the TRINAT data,
using different combinations of kinematic variables, 
as discussed in Refs.~\cite{behrenam,gorelov}.
An example is shown in Figure~\ref{behrfig2}, where the change in the angular
distribution due to $s$ or to $a$ is shown, with the result that 
$s$ produces less than a 0.002 change in $a$~\cite{gorelov}.  
The effect of $s$ on the deduced value of 
$a$ depends on the experimental geometry, the experimental observables,
and on the value of $a$ itself. 
In pure Fermi decays, the null in the $\beta$-$\nu$ angular distribution is helpful to tell the difference between
the effects from $s$ and changes in $a$. The effect of $s$   
will be more strongly correlated with $a$ 
in experiments that solely consider the recoil energy spectrum.

\subsubsection{Beta-neutrino correlation summary}
\begin{figure} \begin{center}
\resizebox{0.45\textwidth}{!}{%
\includegraphics{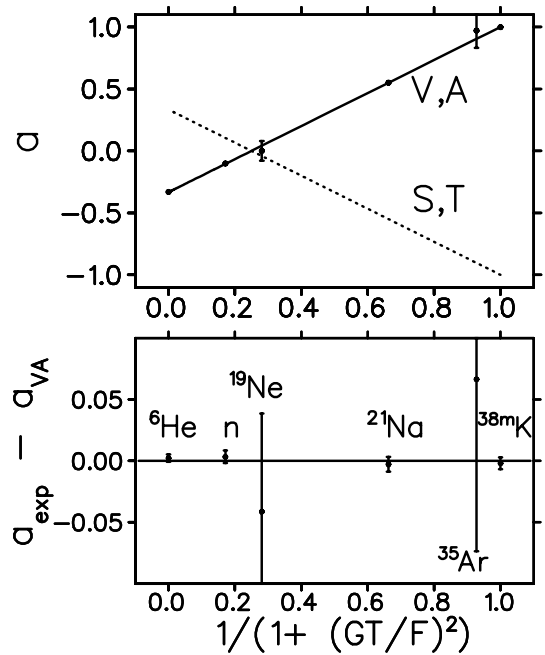}
}
\end{center} 
\caption{
Present status of constraints on non-V,A interactions 
from measurements of the $\beta$-$\nu$ angular correlation coefficient $a$,
updated 
from~\cite{comminsandbucksbaum} (and as first plotted in~\cite{allen35}).
The trap experiments in 
$^{21}$Na~\cite{vetterprc} and $^{38{\rm m}}$K~\cite{gorelov} are shown, 
along with
previous $^{6}$He~\cite{gluck}, n~\cite{stratowa}, 
$^{19}$Ne~\cite{allen35},
and $^{35}$Ar~\cite{allen35} measurements by other techniques.
``GT'' and ``F'' are the Gamow-Teller and Fermi matrix elements, so  
the x-axis variable
is unity for pure Fermi decay and zero for Gamow-Teller decay. The two
trap-based $\beta$-$\nu$ correlation results show the utility of constraints 
with large Fermi components.
}
\label{behrfig5}
\end{figure}

Figure~\ref{behrfig5} summarizes the contribution of $\beta$-$\nu$
correlation measurements, including the trap work in 
$^{21}$Na and $^{38{\rm m}}$K,
to our knowledge of the Lorentz structure of the weak 
interaction. On the horizontal axis is plotted a variable showing the 
degree of Fermi versus Gamow-Teller strengths. The solid line shows the
prediction of V and A interactions. Note that the relative sign between
V and A is not determined, as the $\beta$-$\nu$ correlation is not sensitive
to parity violation.
The dashed line shows the prediction of a pure S,T theory. 
The history of this plot is quite interesting, as in the
late 1950's and early 1960's there were conflicting experimental results in
the $^{6}$He $\beta$-$\nu$ measurement.
Respected yet colorful theorists  
favoring $V-A$ from their conserved vector current hypothesis
suggested that those experiments which were in disagreement must be 
wrong~\cite{feynman}.
The eventual accepted measurement of $a$ in $^{6}$He~\cite{gluck}, together 
with the other measurements of Figure~\ref{behrfig5},
produces tight constraints in agreement with the interaction
being purely V and A.

It is important to recognize that the nuclear structure corrections at
this level are minimal and well under control. The pure Fermi case,
$^{38{\rm m}}$K, is one of the well-characterized isobaric analog 
superallowed $ft$ 
cases. The helicity argument to derive the 
$\beta$-$\nu$ correlation given above relies only
on angular momentum conservation, 
and isospin mixing does not change the Standard Model
prediction of $a=1$. 2nd-order forbidden terms where the leptons carry off
orbital angular momentum are suppressed to less than 10$^{-6}$.
Radiative corrections produce real photons which,
if undetected, perturb the momenta and produce a correction of $\approx$0.002
(corrected for in the Monte Carlo used in~\cite{gorelov}). Recoil order
corrections enter at 3$\times$10$^{-4}$~\cite{holstein} and are independent
of nuclear structure. In an upgraded experiment, the TRIUMF group hopes
to achieve 0.001 accuracy~\cite{trinatexp}, 
so the smallness of the theory corrections is important.

We mentioned in Section~\ref{decayexperiments} that in mixed Gamow-Teller/Fermi
transitions, higher-order corrections within the SM are given by CVC. 
The case of $^{6}$He, while not such an isobaric analog decay, is also
very favorable in terms of nuclear structure, because the higher-order
corrections in $\beta$ decay theory are either known or small.
The recoil-order weak magnetism can be related to experimentally
known  M1 $\gamma$-ray decay by
the CVC hypothesis, and although the first-class induced
current $d$ depends on nuclear structure, 
$d$ is very small in this case because of accidentally favorable
structure of the $A=6$ nuclei~\cite{calapricehe6}. The Paul trap measurement
mentioned in the introduction is in $^{6}$He~\cite{ganil}, 
and the Argonne group is considering
a $\beta$-$\nu$ correlation experiment in $^6$He using a dipole force 
trap~\cite{lu3}.

\subsection{Recoil momenta from shakeoff electron coincidences}

The Berkeley group's technique of using the
atomic shakeoff electrons as a time-of-flight trigger~\cite{scielzoelectron}
has other possible experimental applications.

There have been a number of estimates of the shakeoff electron kinetic
energies, which are thought to be approximately twice their atomic
binding energy. They can therefore be collected in the same electric field
that collects the daughter ions into an MCP, and efficiencies of $\sim$ 50\%
can be attained. This enables a variety of high-statistics experiments.
Experiments involving polarized nuclei are outlined in 
Section~\ref{sec:polarized}. 
Here we describe searches for exotic particles in the recoil momentum
spectrum. 

\subsubsection{Sterile neutrino admixtures}
\label{sterile}

First we show the limitations of 3-body decays in missing mass searches.

\begin{figure*} \begin{center}
%\resizebox{0.75\textwidth}{!}
%{%
  \includegraphics[angle=90,width=0.75\linewidth]{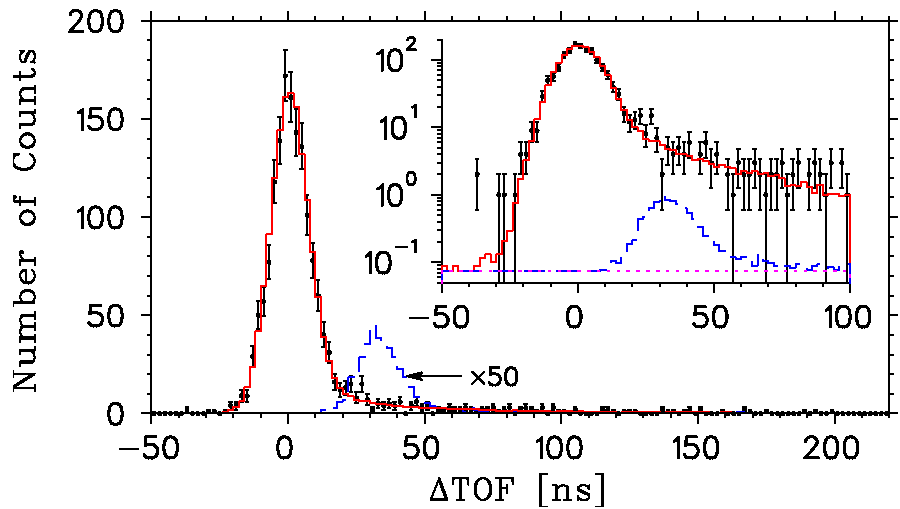}
  \end{center} 
%}
%\vspace*{5cm}       % Give the correct figure height in cm
\caption{
Search in $^{38{\rm m}}$K decay 
for a massive sterile neutrino having admixture with the electron neutrino.
The experimental TOF spectrum is referenced to a simulation of recoils
due to a zero-mass $\nu$.
A simulated 1 MeV $\nu$ (with admixture 50x larger than the experimental
limit) makes the peak shown at delayed TOF, because it has lower 
momentum~\cite{trinczek}.  
}
\label{trinczekfig}       % Give a unique label
\end{figure*}

TRINAT, using the neutral recoils  
from $^{38{\rm m}}$K decay (see Figure~\ref{fig:betanu}),
searched for admixtures of $0.7-3.5$ MeV $\nu$'s with the
electron $\nu$~\cite{trinczek}. 
The results are listed by the Particle Data Group~\cite{pdg}. 
The existence of such $\nu$'s could alter 
astrophysical observables~\cite{abazajian}\cite{dolgov}, and
they can be produced in models with extra dimensions~\cite{mclaughlin}.
The kinematic coincidences effectively reduce the 3-body kinematics to
2-body, and allow a search for peaks in a TOF spectrum instead of the
more conventional search for kinks in continuous $\beta$ 
spectra~\cite{deutsch}. 
The admixture upper limits are as small as $4 \times 10^{-3}$, and
are the most stringent for 
$\nu$'s (as opposed to ${\bar \nu}$'s) in this mass range, 
although there are stronger indirect limits from other
experiments.
Typical results are shown in Figure~\ref{trinczekfig}. 

The 3-body
reconstruction is marred by the $\beta$ detector 
energy response tail seen in Figure~\ref{behrfig2}, 
producing the smooth background seen in Figure~\ref{trinczekfig}.
Though this technique is an improvement over searching for kinks in 
$\beta$ spectra, the sensitivity is limited by statistical fluctuations
in the background and improves only with the square root of the counting time.

Two-body electron capture decay could provide a much cleaner method, and
has the promise of improving existing limits by orders of magnitude.
First we consider a simpler experiment in decay of nuclear isomers.

\subsubsection{ Searches for exotic particles in isomer decay}

\begin{figure*} \begin{center}
%\resizebox{0.75\textwidth}{!}
\includegraphics[angle=90,width=2.4in]{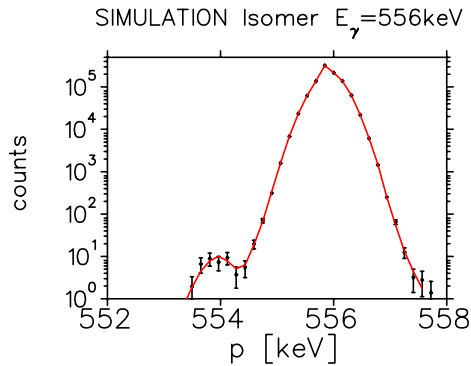}
\end{center} 
%%\vspace*{5cm}       % Give the correct figure height in cm
\caption{
Simulation of the momentum spectrum from the decay of $^{86m}$Rb. The
smaller peak, from a hypothetical particle with mass 50 keV emitted with
branch 4$\times$10$^{-5}$, has lower momentum than the peak from $\gamma$-ray
emission.
}
\label{trinat-galaxycenter}       % Give a unique label
\end{figure*}

TRINAT has begun measurements of 
the momentum of monoenergetic recoils from isomer
$\gamma$ decay.
This makes it possible to search for massive
particles emitted by the nuclear transition instead of $\gamma$-rays.
The recoiling nucleus would have lower 
momentum $p_x = \sqrt{E_\gamma^2 - m_x^2}$, producing a lower-momentum
peak (see Figure~\ref{trinat-galaxycenter}). 
This method does not rely on any information about the
interaction of the particles in any detector, and is independent of
the lifetime of the particle. Given produced yields from ISAC, 
sensitivity to decay branching ratios of $\sim$ 10$^{-6}$ per day of counting
for masses between 20 keV and 800 keV
could be achieved using different Rb and Cs isotopes. Such an experiment 
would utilize
high-momentum resolution spectrometer techniques developed for atomic 
physics experiments in the last decade~\cite{ullrich}, such as TOF drift
spaces and electrostatic lenses to make momentum resolution less dependent
on cloud size.

Angular momentum selection rules favor production of spin and parity
I$^{\pi}$=0$^-$ particles in transitions with magnetic multipolarity, 
and 0$^+$ particles in electric multipole transitions. 
In principle the isomers could also be spin-polarized, and the measured
angular distribution of the recoils would then 
determine the multipolarity of the
emitted particle.

There are  a --- perhaps surprising --- number of phenomenological motivations for  
such ``signature-based searches''. Although the mass range would seemingly
have been explored long ago, potentially there is 
sensitivity to very small couplings that are otherwise difficult to constrain.
These include light 0$^+$ particles 
associated with the annihilation radiation at
the center of the Galaxy~\cite{beacom}, 
0$^-$ particles with smaller couplings than the conventional axion that
could still explain the strong CP problem~\cite{berezhiani}, 
and 0$^-$ particles from
a different global U(1) symmetry that would explain the size of the
$\mu$ parameter in SUSY~\cite{hall}.
Such an experiment is proceeding at TRIUMF in the decay of the 556 keV
isomer in $^{86{\rm m}}$Rb.
Sensitivity at the 10$^{-6}$ level will need to be reached to be competitive with
other, more conventional experiments in this field~\cite{derbin,minowa}.

\subsubsection{Sterile neutrinos in electron capture decay}
A goal is to extend these measurements to electron capture decay and search
for sterile neutrinos with an admixture with the electron neutrino. 
A 1-10 keV mass $\nu_x$ with $\nu_e$ admixtures of
sin$^2$(2$\theta$) $\sim$10$^{-8}$
would be a dark matter candidate~\cite{dodelson,abazajian1}
and have other astrophysical implications~\cite{kusenkopulsar,kusenko2}. 
A theoretical framework for such 
neutrinos
also uses them to moderate inflation 
and produce the baryon asymmetry of the universe~\cite{shaposhnikov1}.

Possible experimental cases include $^{131}$Cs, $^{82}$Sr, and $^7$Be.
A first-generation experiment in $^{131}$Cs
could reach statistical sensitivity to 
admixtures of sin$^2$(2$\theta$) $\approx$ 10$^{-5}$ 
for $m_\nu \approx$ 50-300 keV. 
This would be two orders of magnitude
better than present experiments,  and would be useful to 
constrain scenarios with low post-inflation 
reheating temperatures that produce fewer sterile $\nu$'s~\cite{gelmini}.

Improving the mass resolution to $\sim$ 10 keV 
would require the simultaneous detection and
measurement of all Auger electrons, which carry off several percent of the
momentum of the initial neutrino. Gating on X-ray transitions that select
higher-lying atomic states with fewer Auger electrons could work, though
that would push the energy resolution, efficiency, and time resolution of 
X-ray detectors.

In more standard cosmological scenarios, there are stringent constraints
on the admixtures of these neutrinos, as they tend to overclose the universe.
These $\nu$'s have a two-body decay mode into $\nu_e$ + $\gamma$, and
direct searches for keV X-ray lines have set stringent constraints as 
well~\cite{boyarsky}.
Nevertheless, theorists have suggested experiments in $\beta$-recoil 
coincidences in tritium decay that could reach this sensitivity~\cite{bezrukov}.

\subsubsection{Efficient magnetostatic loading techniques and tritium}
An efficient loading technique which does not use laser cooling 
and can work on a very wide variety of
neutral species has recently been demonstrated. 
Magnetostatic pulses switched
with microsecond periods have been used to slow and trap 
Rb~\cite{raizen} and hydrogen~\cite{hogan} by different groups.
The Rb atoms were then optically pumped to 
an atomic state with different g-factor to escape the magnetostatic trap 
and end up in a dipole force trap, a much better environment for precision
spectroscopy. 
The proponents intend to trap tritium and
measure the electron $\nu$ mass directly by $\beta$-recoil 
coincidences~\cite{raizen}.

\subsection{ $\beta$-decay experiments with polarized nuclei}
\label{sec:polarized}

There are a number of possible correlations to measure if the nuclei
are polarized (see Eq.~(\ref{eq-W})). Traps can provide
high and well-quantified nuclear polarization. When combined with the
detection of nuclear recoils, new and unique correlations can be measured.

\subsubsection{Physics motivations for experiments with polarized nuclei}
The Standard Model electroweak 
bosons couple only to left-handed neutrinos, 
and hence the current is termed V-A.
Experiments with polarized nuclei in
which the polarization can be known atomically can search
for the presence of a right-handed $\nu$. 
Much of the two-parameter space in the simplest ``manifest''
left-right symmetric 
models has been excluded by proton-antiproton collider experiments
and by superallowed $ft$ values~\cite{d0,towner}.
Indirect limits from the K$_{\rm L}$- K$_{\rm S}$ mass difference also
strongly constrain left-right models,
although these limits have some model dependence; e.g.,                   
reasonable simplifying assumptions must
be made about the complicated Higgs sector in left-right
models~\cite{herczeg,langacker}.

However, in more 
complicated non-manifest left-right models,
beta decay measurements with polarized nuclei are still competitive
~\cite{thomas,severijns}. For an example of a specific model, 
we mentioned above the semileptonic scalar and
tensor interactions that can be produced in SUSY and produce
observables at 0.001 level~\cite{profumo}.

\paragraph{ 2nd-class currents}
The leptons and quarks come in weak isospin doublets, which provides
cancellations necessary for the theory to be renormalizable~\cite{weinberg}. 
When QCD 
dresses the quarks into the non-Dirac particle nucleons, 
the isospin symmetry produces a number 
of constraints on the resulting possible currents. 
The absence of isospin-violating ``2nd-class''~\cite{wilkinson} currents 
can be tested in both polarized and
unpolarized observables in isospin-mirror mixed Fermi/GT decays, 
like $^{21}$Na and $^{37}$K. 

\subsubsection{Experiments with polarized atoms in traps}
The Berkeley group's publication of $a$ also measured  
weak magnetism in agreement with the Standard Model~\cite{scielzolbl}, 
i.e. consistent with
no 2nd-class currents,  
although the value achieved is not yet competitive. 
Berkeley has measured precision hyperfine splittings
in $^{21}$Na using optical hyperfine pumping and microwave
transitions~\cite{rowe}; these techniques are applicable
to $\beta$-decay experiments with polarized nuclei.

After demonstrating polarization of  
$^{82}$Rb ($t_{1/2}=76$ s) in a magnetostatic TOP trap~\cite{vieira}, 
the Los Alamos trapping group has since loaded
a dipole force trap with 10$^4$ atoms of $^{82}$Rb. 
They have observed an unusual
spontaneous polarization phenomenon in this trap 
that has been observed before in 
dense gases, and this would be highly useful for $\beta$-decay 
experiments~\cite{feldman}.

\begin{figure*} \begin{center}
  \includegraphics[angle=0,width=0.75\linewidth]{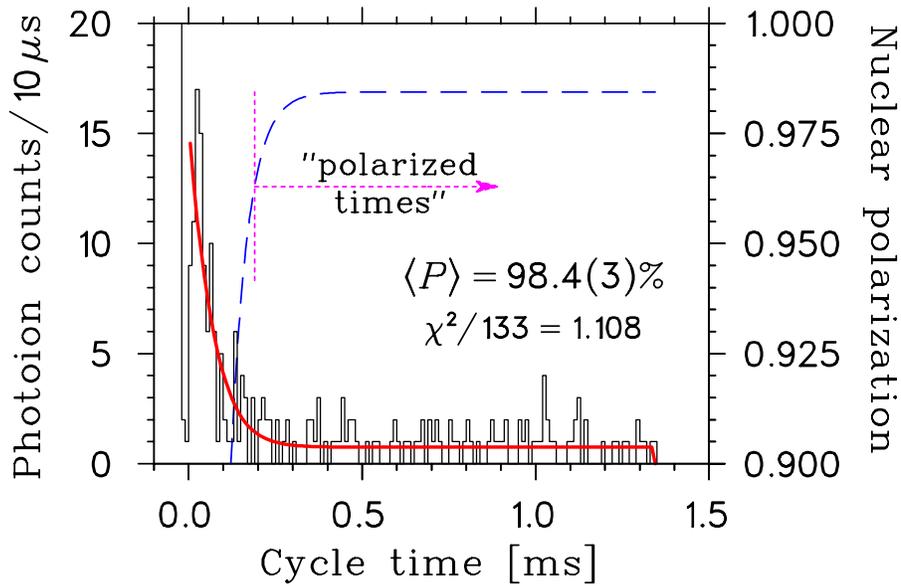}
  \end{center} 
\caption{
The near-vanishing of the fluorescence from the optical pumping of 
$^{37}$K atoms,
which both produces and non-destructively measures the polarization of the 
atoms and nuclei. The polarization (dashed line, right side axis) can be measured continuously for the
same nuclei that decay~\cite{melconian} (see text). 
}
\label{fig-op}       % Give a unique label
\end{figure*}

TRINAT 
has begun experiments with polarized $^{37}$K by turning off the MOT
and optically pumping~\cite{dezafra} the expanding cloud. Circularly polarized 
laser light shines on the atoms (the `D1' beam in Figure~\ref{behrfig1}).
The atoms are excited to states with higher (or lower) angular momentum 
projection, then
decay randomly back to different angular momentum projections.
The state population undergoes a biased random walk, which eventually puts
all the atoms into the ground state with highest (or lowest) angular momentum.
If the excited atomic state has the same total angular momentum as the ground
state (e.g. in alkalis, a S$_{1/2}$ $\rightarrow$ P$_{1/2}$ transition),
then after they are fully polarized, the atoms stop absorbing light.
Using this technique, 
nuclear vector polarizations
of 97$\pm$1\% have been measured by the vanishing of fluorescence in 
S$_{1/2}$ to P$_{1/2}$ optical pumping as the $^{37}$K atoms are polarized 
(see Figure~\ref{fig-op}). 
An advantage of this technique is that the polarization of the same atoms
that decay is continuously measured in a way that does not perturb the 
polarization.
The neutrino asymmetry $B_{\nu}$ of $^{37}$K has been measured to
be $-0.755 \pm 0.020 \pm 0.013$, consistent with the Standard Model value with
3\% error~\cite{melconian}.
This is the first measurement of
a neutrino asymmetry besides that of the neutron.

In these experiments, the atom cloud position and size are measured by 
photoionizing a small fraction of the atoms with a pulsed laser. The
photoions are then accelerated and collected with the same apparatus that
detects the $\beta$-decay recoils, making a 3-dimensional image of the cloud.
This is critical to test for different cloud position as a function of 
polarization state when the sign of the optical pumping is flipped, and is
also critical for the absolute atom location for $\beta$-$\nu$ 
correlations~\cite{melconian,gorelov}.

An additional novel observable is made possible by combining the polarized
nuclei with the detection of the nuclear recoils.  
The spin asymmetry of slow-going recoils (i.e., back-to-back
$\beta$-$\nu$ emission) vanishes in mixed Fermi/Gamow-Teller decays.
This fact is independent
of the Fermi/Gamow-Teller matrix element ratio, so it is independent of
the degree of isospin mixing and the value of V$_{ud}$. Adequate statistics
are difficult to obtain, but the observable is being measured at TRIUMF in
the $^{37}$K experiments. 

Because of the ease of achieving high efficiency of recoil detection and
characterizing the atom cloud pointlike source, 
it is natural in polarized $\beta$-$\nu$ coincidence measurements to 
consider the coefficient $D$ of the time-reversal violating correlation
from Eq.~(\ref{eq-W}), 
$\hat{I} \cdot (\hat{p_{\beta}} \times \hat{p_{\nu}})$.  
Experiments have measured $D \leq 10^{-3}$ using distributed sources  
in $^{19}$Ne~\cite{hallin} and the neutron~\cite{soldner}. 
This observable was
immediately proposed after the discovery of parity violation~\cite{jtw}. 
Experiments in traps have been considered at TRIUMF, Berkeley, and KVI.

\paragraph{Spin asymmetry of recoils: search for tensor interactions}
When parity violation was discovered, a large number of beta decay observables
were suggested in the literature.
Treiman noticed that 
the recoiling daughter nuclei from the $\beta$ decay of polarized
nuclei have average
spin asymmetry 
$A_{\rm recoil}$ $\approx$ 5/8
($A_{\beta}$+$B_{\nu}$).
This vanishes in the allowed approximation for pure Gamow-Teller decays in
the Standard Model,
making it a sensitive probe of new interactions~\cite{treiman}.
It is a very attractive experimental observable, because knowledge of
the nuclear polarization at the 1 to 10\% level is sufficient to be competitive.

Right-handed vector currents do not contribute, because they also cancel
in the sum ($A_{\beta}$+$B_{\nu}$). 
This leaves $A_{\rm recoil}$ uniquely sensitive to  
lepton-quark tensor interactions.
A renormalizable interaction that Lorentz-transforms like a tensor can be
generated by the exchange of spin-0 leptoquarks~\cite{herczeg}.

Using the detection of shakeoff electrons to determine the recoil TOF and
momentum, 
TRINAT has measured the recoil asymmetry with respect to the nuclear spin 
in $^{80}$Rb, with
result $A_{\rm recoil}$= 0.015 $\pm$ 0.029 (stat) $\pm$ 0.019 (syst).
The systematic error is limited by knowledge of 1st-order recoil corrections
in this non-analog Gamow-Teller transition, which can be constrained by 
the dependence of $A_{\rm recoil}$ on recoil momentum.
This result
puts limits on a product of left-handed and right-handed tensor interactions
that are complementary to the best $^6$He $\beta$-$\nu$ correlation
experiment~\cite{pitcairn}.

\subsubsection{ Circularly polarized dipole force trap}

One type of neutral atom trap only confines fully polarized atoms.
A circularly polarized far-off resonant dipole force trap 
(CFORT) for Rb was efficiently loaded and demonstrated 
to achieve very high spin polarization at JILA in Boulder~\cite{duerr}. 
A dipole force trap from a diffraction-limited focused beam 
ordinarily traps atoms if it is tuned to the red of resonance, and
expels them if tuned to the blue. If linearly polarized 
light is tuned just to the blue of
the S$_{1/2}$$\rightarrow$$P_{1/2}$ (D1) resonance, 
it repels all the atoms. 
However, if the atoms are fully polarized, 
the coupling of circularly polarized light to the D1 transition
vanishes. The same coupling coefficients
apply as for real absorption, and the atoms already have maximum 
angular momentum and cannot absorb more. 
The light is still red-detuned with respect
to the D2 transition, so the fully polarized substate, and only
that substate, is trapped. 
The quantization axis is defined by the laser light direction. 
This trap is not limited by imperfect circular polarization, which merely
makes the trap shallower (the spoiling of the polarization by stimulated
Raman transitions is a negligibly small effect). 
TRINAT has worked on developing this trap 
in $^{39}$K~\cite{prime} for use in $^{37}$K $\beta$
decay experiments.

\section{Weak interaction atomic physics}

The traps also offer bright sources for Doppler-free spectroscopy, and 
precision measurements could measure
the strength of weak neutral nucleon-nucleon
and electron-nucleon interactions.

In broadest terms, higher-$Z$ atoms are more sensitive to possible new 
short-ranged interactions between leptons and quarks, because the
electron wavefunction overlap with the nucleus is larger. For atomic
parity violation the effects scale like $Z^2 N$ with additional relativistic
enhancement, anapole moments scale like $Z^{8/3} A^{2/3}$, 
and there is similar scaling for electric dipole moment effects. In case of parity violation experiments, detailed knowledge of the atomic structure is necessary to extract the weak interaction physics from the measurement. Currently, only alkali atoms are sufficiently well understood theoretically. The combined requirements of high $Z$ and alkali structure essentially single out francium as the best candidate for an atomic parity violation experiment in that region. EDM research is still in the `discovery phase', where the unambiguous identification is the primary goal; correspondingly, atomic structure knowledge is less relevant, and heavy stable elements such as mercury have played a dominant role. Nevertheless, EDMs are predicted to be significantly enhanced in the presence of nuclear octupole deformation, making certain radon and radium isotopes very interesting candidates for experiments. These considerations lead, rather accidentally, to the use of unstable isotopes for both parity-violation and EDM experiments; i.e. unlike in $\beta$-decay measurements, the radioactivity is not essential, but an unavoidable result of choosing optimum atomic and nuclear properties.

Even with the availability of relatively
copious amounts of the necessary isotopes from the present generation of
radioactive beam facilities, such as ISAC at TRIUMF and ISOLDE at CERN, the number of atoms available for spectroscopy is orders of magnitude lower than for experiments with stable isotopes in beams or vapor cells. However, soon after the invention of laser trapping and cooling,
it was realized that these new techniques could make up for this shortfall.

Experiments in this direction have been pursued at 
Stony Brook, where precision techniques were developed within a MOT 
environment to measure lifetimes and hyperfine splittings of 
several states. A review can be found in~\cite{gomezreview}. 
Several facilities plan work
with radioactive atom traps, including Argonne 
National Lab, KVI 
Groningen, Legnaro, RCNP Osaka, and TRIUMF. 

\subsection{Searches for permanent electric dipole moments of electrons and nuclei}

The existence of permanent electric dipole moments would violate time reversal symmetry (for reviews see~\cite{fortson,pospelovritz}).
The CPT theorem holds for locally Lorentz-invariant quantum field theories.
Then CP violation implies time-reversal violation and vice versa. 
CP violation was observed in K meson decay in the 1960's and more recently
in B meson decay. Most observables are consistent by the Wolfenstein 
parameterization of the CKM matrix phase. 
Electric dipole moments of the electron and the neutron are predicted to
be very small within this Standard Model CP violation mechanism, and their
existence at forseeable accuracy would imply non-Standard Model physics.
The CP violation in the Standard Model is not enough to generate the 
baryon asymmetry of the universe in the method outlined by Sakharov~\cite{sakharov}.

An electric dipole moment of the electron would manifest itself in the case of non-vanishing electronic angular momentum $J \neq 0$
as an atomic electric dipole moment. 
Although the effects are
suppressed by the rearrangement of charge, when relativity is taken
into account there remains an atomic electric dipole moment, and the effects are enhanced in heavier atoms. 

A number of time-reversal violating effects can produce a nuclear 
`Schiff moment'  in $J=0$ atoms. 
These include an electric dipole moment of the neutron or
proton (or their constituent quarks) and time-reversal violating 
interactions. The nuclear Schiff moments are thought to be enhanced  
by octupole deformation~\cite{auerbach}, which is a well-established nuclear phenomenon.
There are experiments underway to take
advantage of this effect. 

\subsubsection{EDMs with radioactives in traps}

Radium has now been trapped at Argonne National Lab~\cite{guest2}.
The difficulty is great, so it is worth discussing some 
technical details here. 

\begin{figure*} \begin{center}
\includegraphics[angle=0,width=0.5\textwidth]{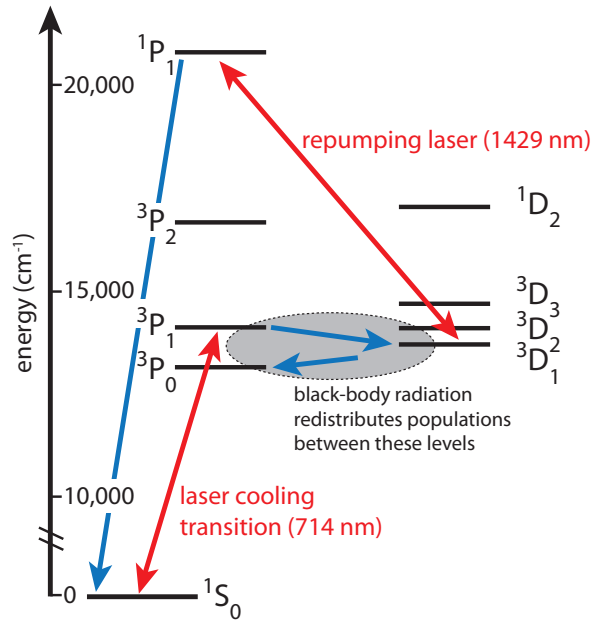} %radium diagram
\end{center}
\caption{
A level diagram for radium with the relevant states for laser trapping. Adapted and simplified from~\cite{guest2}.
}
\label{fig:radium}
\end{figure*}
Radium was known to have a transition that could be pumped at high
power by Ti:Sapph lasers, the  
$^1$S$_0$$\rightarrow$$^3$P$_1$ transition at 714 nm (a level diagram is shown in Figure~\ref{fig:radium}).
This transition appears to be spin forbidden, 
though in this heavy atom the configurations could possibly be mixed enough 
for it to be strong enough for trapping.
Using an atomic beam, 
the Argonne group first measured the $^3$P$_1$ lifetime to be 
420$\pm$20 ns, adequately strong for trapping~\cite{guest1}.  
The $^{225}$Ra is generated from a $^{229}$Th source.
A Zeeman slower was necessary to improve
efficiency, as generally vapor cells do not work for alkaline
earths.

The $^3$P$_1$ state has paths to decay to metastable $^3$D$_2$, $^3$D$_1$, and
$^3$P$_0$ states. A repumping laser at 1429 nm was used to clear the $^3$D$_1$
state, extending trap lifetimes from millseconds to seconds.
Interesting effects from blackbody radiation acting as a repumper were 
observed that cleared the $^3$P$_0$ state~\cite{guest1}.
The plan is to pursue an EDM measurement in the $^{225}$Ra
ground state in a dipole force trap or lattice~\cite{guest1},
taking advantage of well-characterized nuclear octupole enhancement in this
particular isotope. 

In addition to the ground state enhancements from the octupole deformation, 
there are potentially large enhancements in excited states of the radium 
atom. For example, the $^3$D$_2$ excited state,
which is nearly degenerate in energy 
with the $^3$P$_1$ upper state of the trapping
transition, has predicted EDM effects enhanced by nuclear Schiff and
magnetic quadrupole moments by 10$^5$ over mercury, 
and nuclear anapole moment effects enhanced by 10$^3$ over 
cesium~\cite{flambaum}, 
though it remains to be seen whether the lifetime becomes too short when
electric fields are applied to make this a practical system. 
The $^3$D$_1$ state has enhancements over francium 
for the atomic parity-violating E1 transition and electron EDM of 5 and 6,
respectively~\cite{flambaum}.

KVI is building a MOT for barium atoms in preparation for a radium MOT using
the stronger blue-frequency transition, eventually for 
EDM and atomic PV experiments~\cite{wilschut,de}. 
A group at RCNP~\cite{sakemi} has plans for EDM experiments in francium,
and is at the stage of measuring francium production.

%\paragraph{Electron EDM experiments}
Searches for  an  electron EDM is the goal
of a fountain experiment by the group of Gould at LBNL, 
who have measured the scalar dipole
polarizability of cesium~\cite{amini}.
They have published a prototype experiment to
measure the electron EDM with a cesium atomic fountain~\cite{amini2}
including characterization of systematic errors and an outline of upgrades
needed to make it competitive. This group also developed the $^{229}$Th source for  used for $^{221}$Fr trapping at JILA~\cite{lu2}. 
At a radioactive beam facility francium could 
be trapped in similar numbers to stable cesium, and the
higher-$Z$ atom would enhance sensitivity by a factor of 8~\cite{flambaumedm}.

\paragraph{A non-trap EDM experiment on radioactives}
It does not involve a trap, but it is appropriate to  mention  
work in a radon EDM experiment led
by a University of Michigan group. 
The goal is to use the $\gamma$-ray anisotropies or $\beta$ asymmetries
as the Larmor precession probe to measure the EDM of octupole deformed radon
isotopes, which could include 
$^{221}$Rn, $^{223}$Rn, or $^{225}$Rn.
In preparation,
spin-exchange optical pumping on $^{209}$Rn was demonstrated at 
Stony Brook ~\cite{chupp}. 

\subsubsection{Trap efforts for EDMs in stable species}
Laser-cooled atoms and traps have inspired EDM searches in reasonably
high-$Z$ non-radioactive systems. It is beyond the scope of this review to
go beyond a simple mention of the possibilities. 

Ytterbium has been trapped in Kyoto~\cite{honda,honda2}
and Seattle~\cite{maruyama}, 
and groups in those places and at Bangalore~\cite{natarajan} have
proposed EDM experiments in this atom with relatively simple
structure.
Systematics for electron EDM experiments from collisions 
in a optical dipole force trap
were considered originally in~\cite{heinzen}, while
potential systematics for EDMs in
a dipole force trap from light shifts were worked out in detail in~\cite{romalisfortson}.
Spin noise has been investigated in detail experimentally
at Kyoto~\cite{takeuchi} and methods to minimize 
inhomogeneous  broadening in optical dipole traps
were proposed at Seoul~\cite{choi}, with EDM experiments in mind.
There has also been work on 
EDM experiments using optical lattices in Cs~\cite{chin,damopCalgary}.

\subsection{Atomic Parity Violation}

Historically, atomic parity violation (APV) has played an important role. Shortly after the landmark e-D inelastic scattering experiment at SLAC~\cite{prescott78,prescott79} measured the parity violating part of the neutral current weak interaction, APV confirmed these findings at a very different momentum scale. In terms of the electron-quark coupling constants $C_{1u}$ and $C_{1d}$, APV provides constraints nearly perpendicular to those of the SLAC experiment. A sequence of increasingly refined APV experiments throughout the 1980s tightened these constraints to well below those of scattering experiments such as e-D at SLAC and e-carbon at BATES (see e.g. the right panel in Figure~\ref{fig:weinberg}). Until the LEP collaborations published their results, APV even provided a competitive value for $\sin^2{\theta_w}$. This feat is no longer possible in the post-LEP era, but nevertheless low energy experiments still have a key role to play.
For example, when new states are discovered at the LHC, 
it will be important to know their
couplings to the first generation of particles. 
Electrons and muons can be distinguished in the detectors, 
but up/down quark jets cannot be distinguished from jets of other
generations. Atomic parity violation and other low-energy experiments are in a unique position to assist with this question. %ggg
The challenge is to make them sensitive enough, which generally
means part per thousand accuracy. We will describe below experiments in
atomic parity violation in francium that are being 
designed to achieve this accuracy.

The study of weak interactions between nucleons gives unique information
about very short-ranged correlations between them. 
Trapped francium atoms can be used to 
study a parity-violating electromagnetic moment, the anapole moment,
that could provide conclusive
information that these correlations change in nuclear matter.

\subsection{Anapole moments: physics motivation}

The strength of the weak neutral current in nuclear systems remains a
puzzle.
Historically, if the isovector weak meson-nucleon coupling $f_\pi$ had been
larger, weak neutral currents could have been discovered in low-energy
nuclear experiments before Gargamelle's neutrino scattering.

The anapole (`not a pole') moment is a parity-violating 
electromagnetic moment produced by 
the weak nucleon-nucleon interaction. 
It is the result of the chirality acquired by the nucleon current that can be naively decomposed into two parts: a dipole moment, and a toroidal current that generates a magnetic field only in its interior (anapole). It is formally defined as
\begin{equation}
\textbf{a}=-\pi \int d^3r \,r^2 \textbf{J(r)}, \label{definea}
\end{equation}
where $\textbf{J(r)}$ is the electromagnetic current density in the nucleus.
The nuclear anapole comes from a number of effects, though detailed 
calculations suggest it is dominated by 
core polarization by the valence nucleons~\cite{haxton2}. 
This suggestion can be tested by a systematic study of francium isotopes with
paired and unpaired neutrons.

\begin{figure*} \begin{center}
\includegraphics[angle=0,width=2.5in]{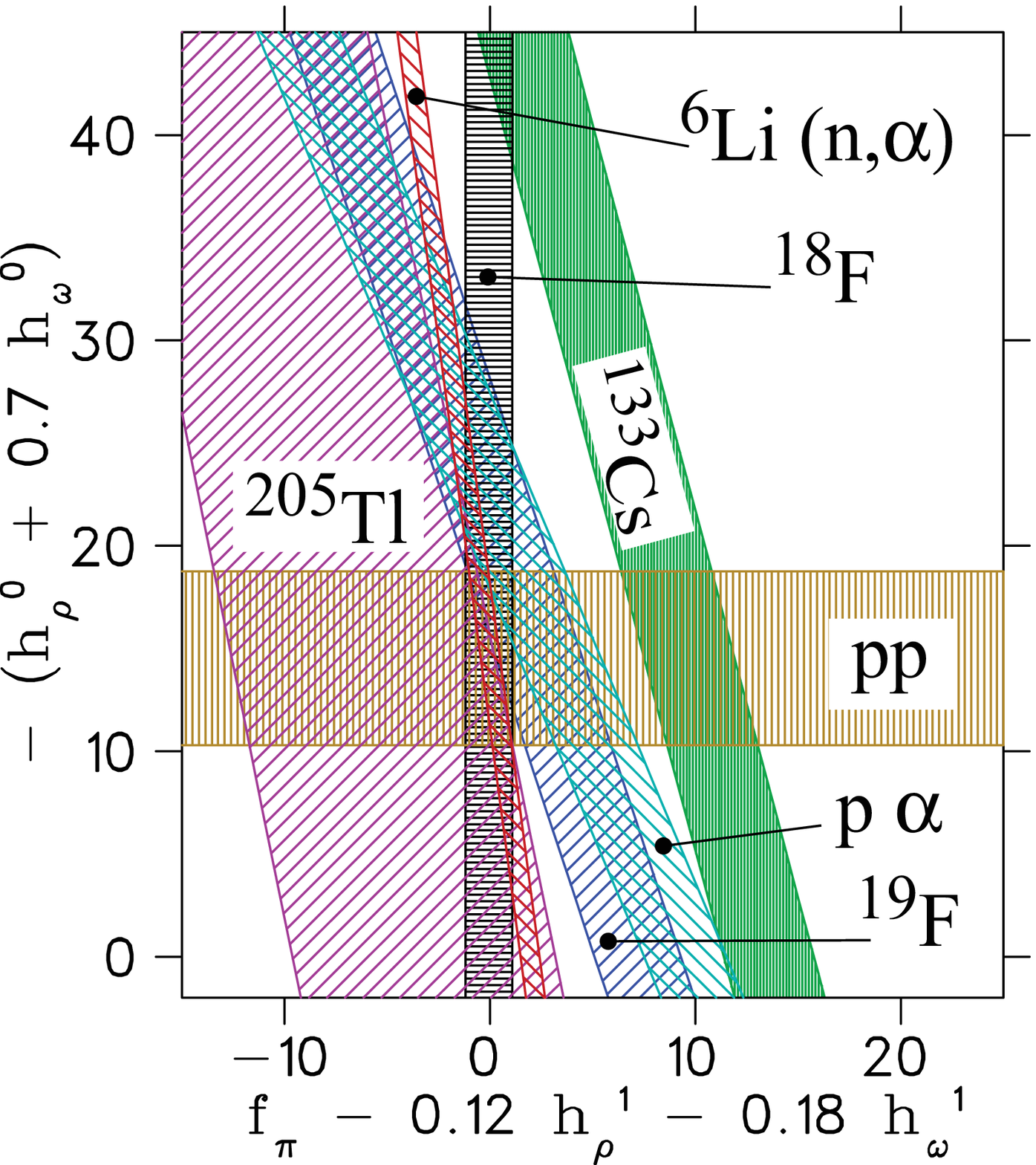}
\includegraphics[angle=0,width=3.0in]{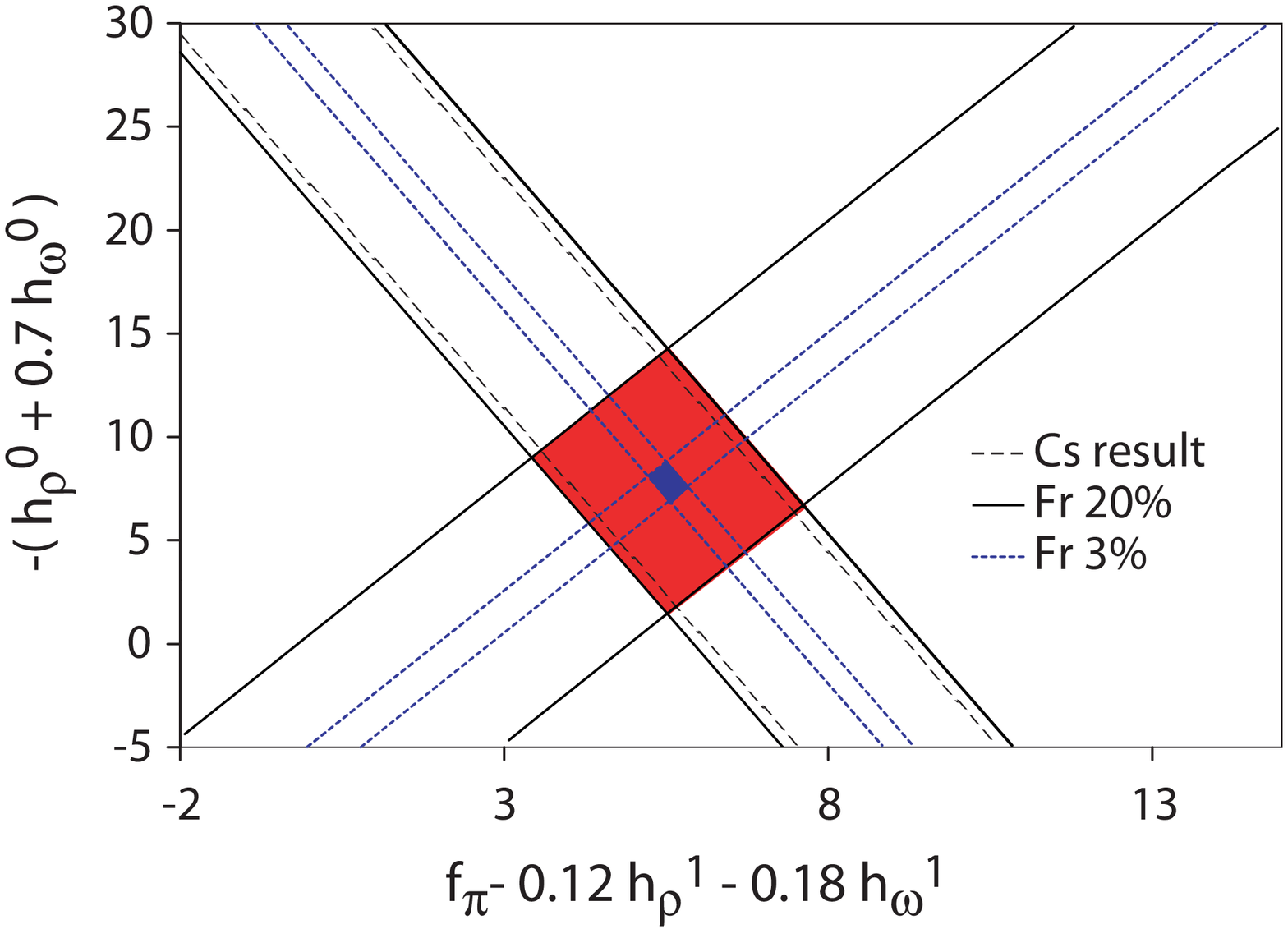}
\end{center}
\caption{
Left: Constraints on isovector and isoscalar weak N-N couplings ($\times$10$^7$) from
measurement of the anapole moment of $^{133}$Cs and natural thallium isotopes,
compared to low-energy nuclear parity violating experiments~\cite{haxton}
including a recent accurate $^{6}$Li(n,alpha) measurement~\cite{vesna}.
Right: {\em Projected} anapole moments of odd-neutron and even-neutron Fr isotopes
would constrain isovector and isoscalar 
weak N-N couplings in the nuclear medium,
if systematic measurements of the odd-even dependence
in several francium isotopes successfully 
show that polarization of the core by the valence neutron is the main effect; courtesy E. Gomez and L.~A. Orozco.
}
\label{fig:haxton}
\end{figure*}

The measurement of the $^{133}$Cs anapole moment
is difficult to reconcile with low-energy nuclear parity-violating
experiments (Figure~\ref{fig:haxton}). 
More cases are needed to understand the basic phenomenon, which is inherently
interesting in itself.
(It could be said that trying to understand nuclear magnetic moments from
two cases would also be a difficult task.)

If the anapole moment values continue to disagree with lighter nuclei and 
few-nucleon systems~\cite{haxton},
this could be due to the modification of the couplings in the nuclear medium~\cite{ramsey}. 
The weak N-N interaction has recently been reformulated as an effective
field theory, and this formalism provides a good framework 
in which to ask whether
the effective couplings derived from few-body systems will be the same
in heavier nuclei~\cite{ramsey}. 

The result could have implications outside of the weak N-N
interaction in another problem which has been reformulated as an effective
field theory: a possible contribution to 
neutrinoless $\beta$$\beta$ decay from exchange of new heavy particles~\cite{prezeauvogel}. 
There are four-quark effective operators that are analogous with those in the
weak N-N interaction, so the degree of renormalization of the weak N-N
interaction 
could be an important guide to their computation. (See the last two pages
of~\cite{ramsey} for a discussion of this issue.)

\subsubsection{Anapole moments: experimental overview}

An anapole experiment in francium to be done at TRIUMF 
is currently in development by the FrPNC collaborators
at the University of Maryland, William and Mary, San Luis Potosi,
Manitoba, and TRIUMF.
The physics method is described in considerable detail
in~\cite{gomez}. We only outline the technique here. 

\begin{figure*} \begin{center}
\includegraphics[angle=0,width=3in]{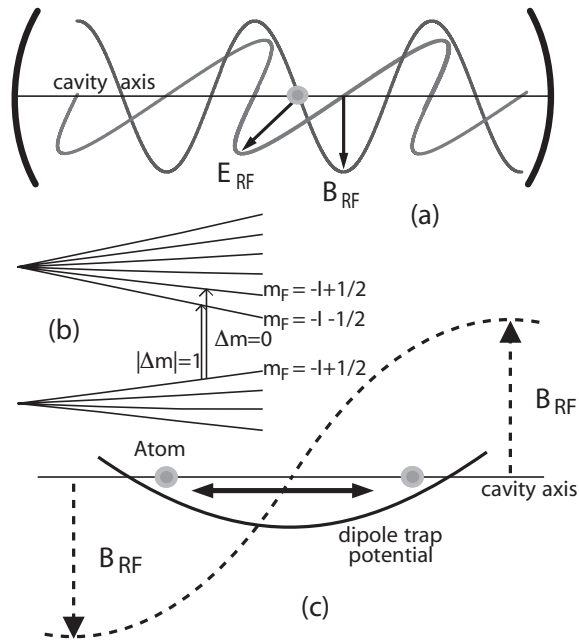}
\end{center} 
\caption{
Schematic indications of the suppression of the allowed M1 transition in the 
anapole experiment. (a) The atoms are placed inside a microwave cavity containing a linearly polarized standing wave. By placing the atom cloud at the node of the magnetic field (and therefore the anti-node of the electric field), the amplitude of the M1 transition can be suppressed. (b) The Zeeman effect due to a static magnetic field applied along the direction of $B_{\rm RF}$ separates the frequency of the M1 transition ($\Delta m = 0$) from that of the  E1 transition ($| \Delta m | = 1$), which is in resonance with the microwave radiation. This  provides additional suppression of the M1. (c) In a finite-sized atom cloud, not all atoms can sit exactly at the node of the magnetic field. However, as long as the trap is precisely centered on the node, the M1 can be ''dynamically suppressed''. Individual atoms slosh back and forth through the trap center. Each time an atom crosses the node, it experiences a phase shift of $\pi$ in the local oscillating magnetic field; if this happens sufficiently often during the coherent microwave excitation of the atom, a strong reduction in the M1 excitation rate can be achieved (taken from~\cite{gomez}, where more details are found).
}
\label{fig:anapoleexp}
\end{figure*}

In the Boulder Cs and the Seattle Tl experiments, the anapole was extracted by determining the difference in the atomic parity violation signal on two different hyperfine transitions ($nF \rightarrow n'F'$ and $nF' \rightarrow n'F$), i.e. taking the difference of two very similar numbers. As a result, the relative error on the anapole measurement is much larger than that of the nuclear-spin independent part. One way of addressing this problem is to measure atomic parity violation on a transition where the nuclear-spin independent part is absent, e.g. within a ground state hyperfine manifold, as was proposed long 
ago~\cite{sandars}.
A PV-induced E1 transition between hyperfine states is driven by microwave 
radiation in a high-finesse cavity (see Figure~\ref{fig:anapoleexp}). 

The M1 between these states is allowed and must be suppressed
by orders of magnitude in contrast to the optical
experiment (see below).
Three simultaneous methods to do this are sketched
broadly in Figure~\ref{fig:anapoleexp}.
Together~\cite{gomez} estimates that the M1
amplitude can be reduced to less than 1\% of the PV E1 amplitude 
(see Figure~\ref{fig:anapoleexp}).

\paragraph{Other efforts: anapole moments}
DeMille at Yale is planning to measure anapole moments by placing
diatomic molecules in a strong magnetic field~\cite{demilleanapole}. A
collaboration in Russia wants to measure the anapole moment in a
potassium cell~\cite{ezhov04}. The Budker group in Berkeley has been pursuing measurements in ytterbium, which has many stable isotopes available~\cite{stalnaker02}, and with the appropriate hyperfine transitions could extract anapole moments. Other suggestions using atomic fountain techniques have
recently appeared in the literature~\cite{bouchiat07}.

\subsection{Atomic parity violation in francium: physics motivations}
\begin{figure*} \begin{center}
\includegraphics[angle=0,width=3.5in]{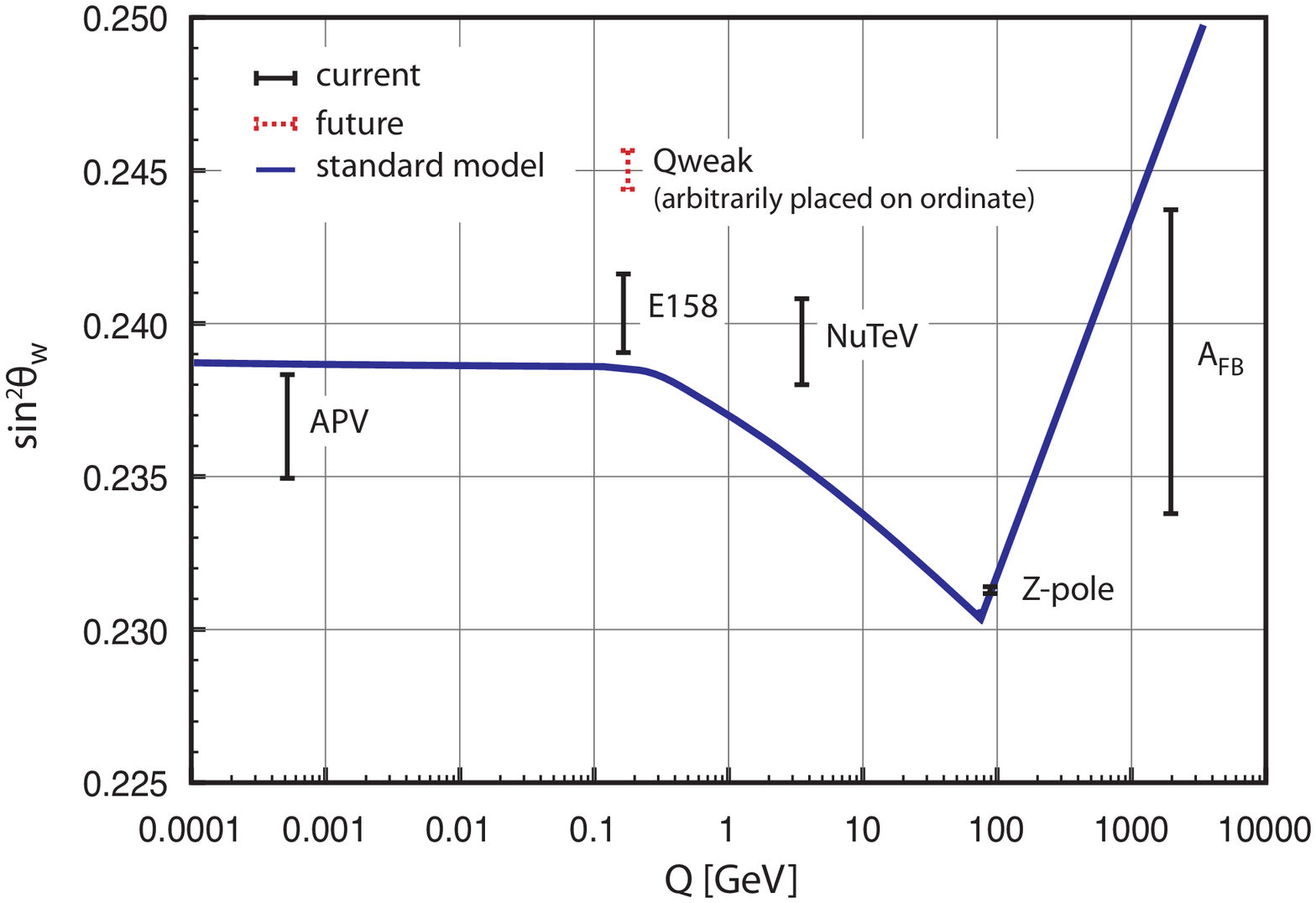}
\includegraphics[angle=0,width=2.6in]{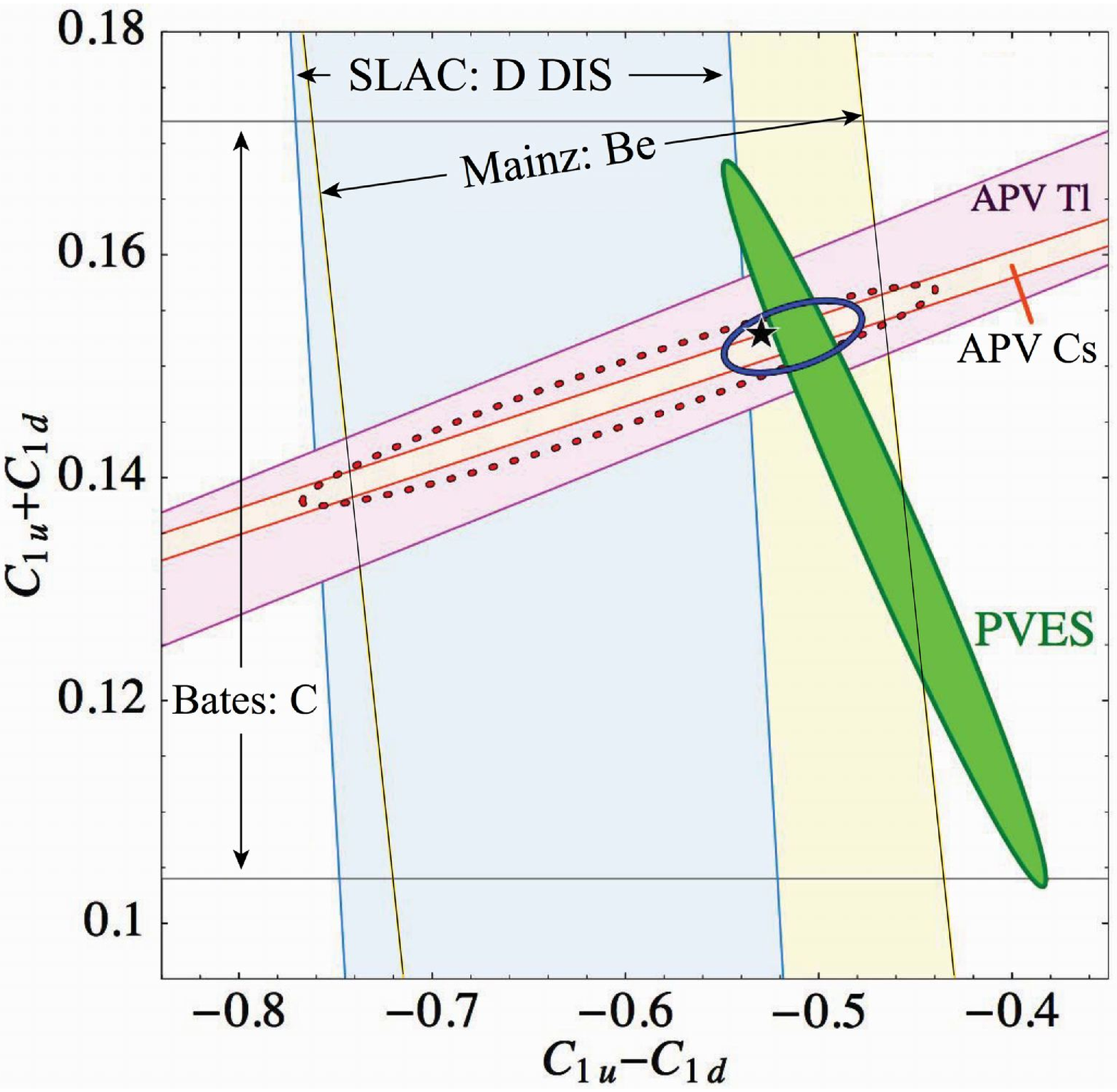}
\end{center}
\caption{
Left:
Measurements of the weak neutral current strength as a function of momentum
transfer ($\theta_w$ is the Weinberg angle). Despite their lower precision, the low-energy experiments 
retain useful sensitivity to exchange of new bosons because they reside on
the tail of the Standard Model Z resonance. Adapted from a figure courtesy J. Erler, see also~\cite{Erler:2004in,ramseysu}. The blue line is the Standard Model prediction.
Right:
Constraints on weak quark couplings from electron scattering and atomic
parity violation from~\cite{young}, showing their complementarity. The blue star denotes the Standard Model prediction.
}
\label{fig:weinberg}
\end{figure*}

Atomic parity violation measures the strength of the weak neutral current at 
very low momentum transfer. 
There are three types of such low-energy weak neutral current measurements
with complementary sensitivity. The atomic weak charge is predominantly 
sensitive to the neutron's weak charge, as the proton weak charge is 
proportional to $1-4\sin^2{\theta_W}$, which accidentally is near zero. 
At Jefferson Lab, the upcoming Qweak electron scattering experiment  on hydrogen 
is sensitive to the proton's weak 
charge. The SLAC E158 Moeller scattering is sensitive to the electron's weak 
charge. 
Different Standard Model extensions then contribute 
differently~\cite{ramseysu}. 
For example, the atomic parity weak charge is relatively insensitive to one-loop order
corrections from all SUSY particles, so its measurement provides a benchmark
for possible departures by the other ``low-energy'' observables. 
As another example, Moeller scattering is purely leptonic and so has no
sensitivity to leptoquarks, so the atomic parity weak charge can then
provide the sensitivity to those.
Figure~\ref{fig:weinberg} (right) from~\cite{young}
shows the present constraints on weak quark couplings 
from parity violating electron scattering and from atomic parity violation.

Figure~\ref{fig:weinberg} shows measurements of the Weinberg 
angle~\cite{ramseysu}. 
The low-energy experiments still have competitive
sensitivity to certain specific Standard Model extensions compared to the LEP 
electroweak measurements--- LEP's precision is better, but the low-energy
experiments seeking terms interfering with the Z exchange can have inherently 
more sensitivity to tree-level exchange because they work on the tail
of the Z resonance. 
It should be stressed that Figure~\ref{fig:weinberg} cannot do justice to the highly complementary nature of the low-energy experiments, as it only plots the sensitivity to one Standard Model parameter, $\sin^2{\theta_W}$. Since Qweak and APV probe different quark combinations and E158 probes leptons, the sensitivities to physics beyond the SM are very different.

An explicit example is given by a recent review on constraints on new
Z' bosons by Langacker~\cite{langacker08}.
Limits on the mass of new Z' bosons in several models 
and their mixing angle with the Standard Model Z are shown in his figure 1 and
his table 4. The mixing angle constraints from `global precision electroweak'
fits are dominated by the LEP measurements at the Z pole, while the mass
constraints come mainly from the low-energy atomic PV and electron
scattering experiments. Those mass limits are at $\approx$ 600 GeV at 
90\% confidence, while
direct searches at the Tevatron (assuming decays into Standard Model particles
only) and at LEP 2 have recently reached better limits of $\approx$ 800 GeV. 
The mass reach
of the low-energy measurements scales roughly with the square root of their
accuracy, so improvements of 2 to 4 in accuracy would again provide 
useful information.

\paragraph{Constraints on parity-violating low-energy physics}
Recently a new scalar particle with mass on the order of a few MeV, along 
with a new exchange boson with slightly greater mass, 
has been invoked to explain a possible excess of 511 keV photons at the 
galactic centre. APV places severe constraints on parity-violating
interactions at low energy, 
so it could immediately be concluded that the new exchange
boson must have purely vector, parity-conserving 
couplings~\cite{bouchiatfayet}. 
This demonstrates the power of the APV measurements to constrain
exotic physics which can surprisingly evade all other constraints.

\subsubsection{Status of atomic parity violation measurements}

The weak interaction in atoms induces a mixing of states of
different parity, observable through APV measurements. Transitions
that were forbidden due to selection rules become allowed through
the presence of the weak interaction. The transition amplitudes
are generally small and an interference method is commonly used to
measure them. A typical observable has the form
\begin{equation}
|A_{PC}+A_{PV}|^2 = |A_{PC}|^2+2Re(A_{PC}A_{PV}^*)+|A_{PV}|^2,
\label{eq:apv_if}
\end{equation}
where $A_{PC}$ and $A_{PV}$ represent the parity conserving and
parity non-conserving amplitudes. The second term on the right
side corresponds to the interference term and can be isolated
because it changes sign under a parity transformation. The last
term is usually negligible.

Most recent and on-going experiments in atomic PV rely on the
large heavy nucleus (large $Z$) enhancement factor proposed by the
Bouchiats (\cite{dunford} is a recent review of PV prospects in hydrogen).
These
experiments follow two main strategies (see recent review by M.-A.
Bouchiat~\cite{bouchiat01}). The first one is optical activity in
an atomic vapor. 
This method
has been applied to reach experimental precision of 2\% in
bismuth, 1.2\% in lead, and 1.2\% in thallium.

The second strategy measures the excitation rate of a highly
forbidden transition. The electric dipole transition between the
$6s$ and $7s$ levels in cesium becomes allowed through the weak
interaction. Interference between this transition and the one
induced by the Stark effect due to the presence of an static
electric field generates a signal proportional to the weak charge.
The best atomic PV measurement to date uses this method to reach
a precision of 0.35\%~\cite{wieman,wood1,wood2} (note a recent announcement of new calculations reducing the theory error further~\cite{derevianko}).

Other methods have been proposed, and some work is already on the
way. 
We have mentioned above Budker's work in optical transitions in 
ytterbium~\cite{stalnaker02}. 
The Bouchiat group in Paris has worked on the
highly forbidden $6s$ to $7s$ electric dipole transition in a
cesium cell, but detects the occurrence of the transition using
stimulated emission rather than fluorescence;
this effort has ended after reaching 2.6\% statistical accuracy~\cite{guena05}. 

Possible advantages for laser-cooled and slowed atomic beams for
APV studies have been considered by the Bouchiat
group~\cite{sanguinetti}.
More recently, Bouchiat has suggested methods to measure anapole moments and electron-nucleon
atomic PV by frequency shifts using fountains and atom
interferometric methods, possibly working on as few at 10$^4$
atoms~\cite{bouchiatfort,bouchiat07}.
These methods would avoid losses from two-photon ionization
discussed below.

There are experimental efforts by
the Fortson group in Seattle using a single barium ion and the
KVI group using a single radium ion~\cite{koerber03,sahoo}.

The group at INFN in Legnaro has trapped $\sim$ 1000 Fr atoms in a MOT, and is considering
atomic physics experiments including atomic parity violation~\cite{rnb7cotina,sanguinetti}. 
They have pioneered a number of innovative loading 
techniques~\cite{atutov2} in stable Rb and are 
in the process of applying these to Fr.
The FrPNC collaboration, in addition to the planned anapole measurement
described above,  is also working towards PV measurements in francium at TRIUMF.

This list is not
intended to encompass all the efforts, but represents some of the
groups interested in PV at present.

%\subsubsection{Considerations for a parity violation experiment in francium}

\begin{figure*} \begin{center}
\includegraphics[angle=0,width=2.5in]{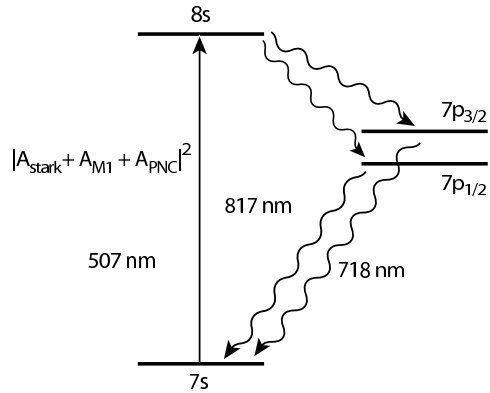}
\end{center} 
\caption{
The most relevant atomic levels for Stark mixing experiments in francium.
}
\label{fig:m1}
\end{figure*}

\subsubsection{Atomic parity violation in francium: further experimental techniques}

So far, there has been no
parity violation measurement in neutral atoms performed
utilizing the new technologies of laser cooling and trapping.  In
order to create a road-map for an experiment, one could assume a
transition rate measurement following closely the technique used
by the Boulder group in cesium~\cite{wood1,wood2}. We start with a Stark
shift to induce a parity conserving amplitude between the $7s$ and
$8s$ levels of francium, and look how this
electromagnetic term will interfere with the weak interaction
amplitude (Eq.~(\ref{eq:apv_if})). This gives rise to a left-right asymmetry with respect to
the system of coordinates defined by the static electric field
{\textbf{E}}, static magnetic field {\textbf{B}}, and the Poynting
vector {\textbf{S}} of the excitation field, such that the
observable is proportional to ${\textbf{B}}\cdot({\textbf{S}}
\times {\textbf{E}})$.

Francium atoms would accumulate in a magneto-optic trap (MOT). Then, after further
cooling to control their velocities,  they would be transferred to
another region where a dipole trap will keep them ready for the
measurement. After being optically pumped to one hyperfine state,
the atoms would be exposed to an intense standing wave mode of 507 nm light resonant with the $7s$ to $8s$ transition, in the presence of a DC electric  field.
Excited atoms will
decay via the $7p$ state
to populate the empty hyperfine state. Optical
pumping techniques allow one to recycle the atom that has
performed the parity non-conserving transition many times
enhancing the probability to detect the signature photon.

\subsubsection{Ramping up to atomic PV: `Forbidden' M1 in atomic francium}

The strength of the `forbidden' M1 in atomic francium is sensitive to
relativistic corrections to many-body perturbation theory~\cite{savukov}.
These effects are useful tests of the atomic theory needed to extract 
weak coupling coefficients from 
atomic parity-violation experiments.
Thus a logical precursor to any optical APV experiment in francium is the spectroscopy of the $7s \rightarrow 8s$ transition (see Figure~\ref{fig:m1}). 
The line is best located by driving the Stark-induced amplitude in a strong electric field (several kV/cm) in a configuration of parallel external field and laser polarization, where the large scalar transition polarizability $\alpha$ provides a (relatively) strong signal.
With crossed field and polarization, the 30 times weaker transitions characterized by the vector transition polarizability $\beta$ then allows to determine the ratio $\alpha/\beta$. Observing the E1-M1 interference by flipping fields similar to the APV procedure, produces intensity modulation at the 1 \% level, about a hundred times larger than the modulation expected in APV.  The quality of this signal will be a crucial indicator for the prospects of observing a $10^{-4}$ modulation to better than 1 \% --- the eventual goal for APV.

\subsubsection{Signal-to-noise ratio for atomic parity violation}

To estimate the requirements for a parity violation
measurement in francium it is good to take the Boulder Cs
experiment as a guide~\cite{wood1,wood2}.
The most important quantity to estimate is
the signal-to-noise ratio since that will determine many of the
requirements of the experiment.
The approach of Stark mixing works as an amplifier in the full
sense of the word, it enlarges the signal, but it also brings
noise. The Stark-induced, parity conserving part $|A_{PC}|^2$ not only dominates the transition strength, it also contributes essentially all of the shot noise to the measurement. The number of excitations in a sample of $N$ atoms is given (up to an angular momentum factor of order one depending on geometry and polarization) by
\begin{equation}
S_{\rm{stark}}=\frac{2}{c \hbar^2 \epsilon_0} I \tau\: (\beta E)^2 N;
\label{stark1}
\end{equation}
the parity violation signal is
\begin{equation}
S_{\rm{PV}} = \frac{2}{c \hbar^2 \epsilon_0} I \tau\: 2 E\beta\: \textrm{Im}(E_{\rm PV}) N,
\end{equation}
where $\beta$ is the vector Stark polarizability, $E$ is the dc
electric field used for the Stark mixing, ${\rm{Im}}(E_{\rm{PV}})$ is the parity violating amplitude, $\tau$ is the lifetime of the upper $s$-state, and $I$ the 
intensity of the excitation source. The polarizabilities are quoted exclusively in atomic units in the literature, and the corresponding value in SI units is obtained by $\beta_{\rm SI} = \beta_{au}/6.06510 \times 10^{40}$. Since the Stark-rate dominates by orders of magnitude, and assuming only shot noise as
the dominant source of noise, the signal to noise ratio achieved
in one second is
\begin{equation}
\frac{S_{\rm PV}}{N_{\rm noise}} = 2 \sqrt{\frac{2\; I \tau N}{c \hbar^2 \epsilon_0}} {\rm Im}(E_{\rm PV}).
\end{equation}
The calculated value from Dzuba \textit{et al.}
~\cite{dzuba06} for ${\rm{Im}}(E_{\rm{PV}})$ of $1.5
\times 10^{-10}$ in atomic units is eighteen times larger than in
cesium.

A serious complication for a trap-based experiment is photoionization in the excited state by the intense 507 nm radiation, which was already discussed in~\cite{Vieira1992}. At intensities of 800 kW/cm$^2$ as used by Wood et al.~\cite{wood1,wood2}, the probability for photoionization per excitation was 10 \%. In a beam experiment, where each atom is used only once, this is not particularly concerning. In a trap scenario, each atoms must be re-used over a time span of up to seconds, and hence, the photoionization rate must be brought down to a compatible level (accidentally, in Fr the situation is worse, as the 507 nm light  can ionize into the continuum from both the $8s$ and the $7p_{3/2}$ states). This can be remedied by reducing the light intensity by a factor of 300,
which will bring the photoionization rate down to about 1 Hz, yielding a 7s-8s excitation rate of 30 Hz per atom. For guidance, we can refer to the Cs experiment which had a $6s-7s$ excitation rate of $10^{10}$ Hz and find that $3 \times 10^8$ trapped atoms lead to the same signal, but the fluorescence modulation upon parity reversals is $2\times10^{-4}$, about an order of magnitude larger. The signal-to-noise is then
\[S/N = 2 \times 10^{-4} \sqrt{30 t N},\]
where {$t$} is the observation time in seconds and {$N$} the number of atoms in the trap. Or, the time to obtain a $S/N$ with a certain excitation rate {$R$} and {$N$} atoms in the trap and an asymmetry {$A$} is
\[t = \frac{(S/N)^2}{A^2 R N}.\]
Based on these purely statistical considerations, a $1.0$ \% APV measurement requires about 2.5 hours using $10^6$ trapped atoms; ten times more atoms would allow a $0.1$ \% test in 25 hours.
Naturally, much more time has to be spent to deal with systematic effects.

\subsubsection{Neutron radius question}
Since the weak charge in atoms stems mostly from the neutrons, %ggg
there is some dependence on the neutron distribution in the nucleus,
a quantity with few reliable experimental probes.
The neutron radius measurement with parity-violating electron scattering 
at Jefferson Lab (`PRex'~\cite{horowitz}) would result in an uncertainty 
on the weak charge in $^{212}$Fr of 0.2\%~\cite{sil}.
Isotopic ratios would need a next generation neutron radius experiment~\cite{sil}, though a recent analysis suggests that when cancellations
in correlated nuclear theory errors are taken into account, new physics
can indeed by extracted by measuring chains of isotopes~\cite{brown},
which also have the potential to remove much of the atomic theory
uncertainty.

Work at Stony Brook investigated the hyperfine anomaly in 
$^{208-212}$Fr~\cite{grossman}. Different atomic wavefunctions have different
overlap with the nucleus, so a changing spatial distribution of nuclear 
magnetism will change the relative hyperfine splittings. For the odd-neutron
isotopes, this effect is sensitive to the spatial wavefunction of the 
valence neutron, in a manner similar to magnetic multipoles in electron
scattering. This effect will be measured in the chain of francium isotopes
in an upcoming experiment at TRIUMF~\cite{pearson}.

\section{ Conclusion}

Neutral atom traps provide unique environments for precision experiments
using radioactive isotopes.
The first trap-based measurements in $\beta$ decay have been completed, and the results are improving constraints on interactions
beyond the Standard Model.
The ability to measure the momentum of the daughter nuclear recoils has produced two of the best $\beta$-$\nu$ correlation experiments.
Adding the ability to reverse the spin of highly polarized atoms
leads to unique observables with the potential to improve parity and time-reversal violation tests
in $\beta$ decay.

Results from francium atomic spectroscopy have long been in evidence.
Plans are proceeding to harness the trapping technologies to measure weak neutral current effects in atoms and nuclei. By storing individual atoms for seconds, a sufficient sample of radioactive atoms can be provided with realistic production rates at radioactive beam facilities. The challenge will be to understand and control systematic errors in an online environment.
Several labs have plans for time-reversal-violating electric dipole moment searches in radium, radon, and francium. Undoubtedly, these efforts will produce exciting new results in coming years.

\ack{This work was supported by 
the Natural Sciences and Engineering Council of Canada and
National Research Council Canada through TRIUMF.
JB thanks innumerable TRINAT collaborators whose work appears
here, in particular O. H\"{a}usser. GG thanks S.~A. Page. GG and JB thank G.~D. Sprouse and L.~A. Orozco.
}\\

\end{document}